\documentclass[12pt]{article}
\usepackage[top=3cm,right=3cm,left=2cm,bottom=3cm]{geometry}

\usepackage{amsmath,amssymb,amsthm}

\numberwithin{equation}{section}

\DeclareMathOperator{\rank}{rank}

\DeclareMathOperator{\dive}{div}
\newcommand{\tr}{tr}

\newtheorem*{prop}{Proposition}

\theoremstyle{definition}

\newcommand{\ba}{\bar{a}}

\newcommand{\bu}{\bar{u}}

\newcommand{\bx}{\bar{x}}

\newcommand{\bt}{\bar{t}}

\newcommand{\ve}{\vec{e}}

\newcommand{\vu}{\vec{u}}

\newcommand{\vm}{\vec{m}}

\newcommand{\vx}{\vec{x}}

\newcommand{\sech}{\mathop{\mathrm{sech}}}

\newcommand{\AB}{\allowbreak}

\newcommand{\ali}[2]{\mathop{\mathfrak{#1}(#2)}\nolimits}

\newcommand{\ad}{\mathop{\mathrm{ad}}\nolimits}

\newcommand{\ADA}[1]{\ifmmode \ad(#1) \else $\ad(#1)$\fi}

\newcommand{\LI}[2]{\ifmmode#2_1,\AB\,\ldots,\,\AB #2_{#1}%

\else$ #2_1,\AB\,\ldots,\,\AB#2_{#1}$\fi}

\newcommand{\sltwo}{\ifmmode \ali{sl}{2} \else $\ali{sl}{2}$\fi}

\newcommand{\bMA}[1]{\[\begin{array}{#1}}

\newcommand{\eMA}{\end{array}\]}

\newcommand{\CA}{{\cal A}}

\newcommand{\NR}{{{\mathbb R}}}

\newcommand{\NZ}{{{\mathbb Z}}}

\def\be{\begin{equation}}

\def\ee{\end{equation}}

\def\p{{\partial}}

\def\tr{{\mathrm{tr}}}

\mathchardef\za="710B  %\alpha

\mathchardef\zb="710C  %\beta

\mathchardef\zg="710D  %\gamma

\mathchardef\zd="710E  %\delta

\mathchardef\ze="710F  %\epsilon

\mathchardef\zz="7110  %\zeta

\mathchardef\zh="7111  %\eta

\mathchardef\zy="7112  %\theta

\mathchardef\zi="7113  %\iota

\mathchardef\zk="7114  %\kappa

\mathchardef\zl="7115  %\lambda

\mathchardef\zm="7116  %\mu

\mathchardef\zn="7117  %\nu

\mathchardef\zx="7118  %\xi

\mathchardef\zp="7119  %\pi

\mathchardef\zr="711A  %\rho

\mathchardef\zs="711B  %\sigma

\mathchardef\zt="711C  %\tau

\mathchardef\zu="711D  %\upsilon

\mathchardef\zf="711E  %\phi

\mathchardef\zq="711F  %\chi

\mathchardef\zc="7120  %\psi

\mathchardef\zw="7121  %\omega

\mathchardef\zG="7000  %\Gamma

\mathchardef\zD="7001  %\Delta

\mathchardef\zY="7002  %\Theta

\mathchardef\zL="7003  %\Lambda

\mathchardef\zX="7004  %\Xi

\mathchardef\zP="7005  %\Pi

\mathchardef\zS="7006  %\Sigma

\mathchardef\zU="7007  %\Upsilon

\mathchardef\zF="7008  %\Phi

\mathchardef\zC="7009  %\Psi

\mathchardef\zW="700A  %\Omega

\mathchardef\ze="7122  %\varepsilon

\mathchardef\zvy="7123  %\vartheta

\mathchardef\zvr="7125 %\varrho

\mathchardef\zvs="7126 %\varsigma

\mathchardef\zvf="7127  %\varphi

\title{Conditional Symmetries and Riemann Invariants for Hyperbolic Systems of PDEs}

\bigskip

\author{
A.M. Grundland
\\
Centre de Recherches Math{\'e}matiques, Universit{\'e} de Montr{\'e}al, \\
C. P. 6128, Succ.\ Centre-ville, Montr{\'e}al, (QC) H3C 3J7, Canada \\
Universit{\'e} du Qu{\'e}bec, Trois-Rivi{\`e}res CP500 (QC) G9A 5H7, Canada \\
email address : grundlan@crm.umontreal.ca
\\\\\\\
B. Huard\\
D{\'e}partement de math{\'e}matiques et de statistique,\\
C.P. 6128, Succc.\ Centre-ville, Montr{\'e}al, (QC) H3C 3J7, Canada \\
email address: huard@dms.umontreal.ca
}
\begin{document}

\thispagestyle{empty}
% \Name{Conditional Symmetries and Riemann Invariants for Hyperbolic Systems of PDEs}
% 
% \Author{A.M. Grundland $^\dag$ and B. Huard $^\ddag$}
% 
% \Address{$^\dag$ Centre de Recherches Math{\'e}matiques, Universit{\'e} de Montr{\'e}al, \\
% C. P. 6128, Succ.\ Centre-ville, Montr{\'e}al, (QC) H3C 3J7, Canada \\
% Universit{\'e} du Qu{\'e}bec, Trois-Rivi{\`e}res CP500 (QC) G9A 5H7, Canada \\
% email address : grundlan@crm.umontreal.ca\\[10pt]
% $^\ddag$ D{\'e}partement de math{\'e}matiques et de statistique,\\
% C.P. 6128, Succc.\ Centre-ville, Montr{\'e}al, (QC) H3C 3J7, Canada \\
% email address : huard@dms.umontreal.ca}

\begin{center}
\text{}\\[9cm]
 {\Large Conditional Symmetries and Riemann Invariants\\for Hyperbolic Systems of PDEs}\\[1cm]

A.M. Grundland 
\footnote{Centre de recherches math\'ematiques, Universit\'e de Montr\'eal, C.P. 6128, Succ. Centre-ville, Montr\'eal, (QC) H3C 3J7, Canada; Universit\'e du Qu\'ebec, Trois-Rivi\`eres CP500 (QC) G9A 5H7, Canada; grundlan@crm.umontreal.ca}
\hspace{2cm} B.Huard
\footnote{D\'epartement de math\'ematiques et de statistique, C.P. 6128, Succ. Centre-ville, Montr\'eal, (QC) H3C 3J7, Canada; huard@dms.umontreal.ca}\\[1cm]

CRM-3226\\
Janvier 2007

\end{center}

\newpage \thispagestyle{empty}

\text{}\\[15mm]
\begin{abstract}
This paper contains an analysis of rank-k
solutions in terms of Riemann invariants, obtained from
interrelations between two concepts, that of the symmetry reduction method
and of the generalized method of characteristics for first order quasilinear hyperbolic
systems of PDEs in many dimensions. A variant of the conditional
symmetry method for obtaining this type of solutions is proposed. A
Lie module of vector fields, which are symmetries of an
overdetermined system defined by the initial system of equations and
certain first order differential constraints, is constructed. It is
shown that this overdetermined system admits rank-k solutions
expressible in terms of Riemann invariants.  Finally, examples of
applications of the proposed approach to the fluid dynamics
equations in (k+1) dimensions are discussed in detail.  Several new soliton-like
solutions (among them kinks, bumps and multiple wave solutions) have been obtained.\\

\bigskip

 \begin{center}
 {\bf R{\'e}sum{\'e}}
 \end{center}
 Dans cet article, nous pr\'esentons une analyse des solutions de rang k
exprim\'ees en termes d'invariants de Riemann obtenues \`a partir
des relations entre les m\'ethodes de r\'eduction par sym\'etries et
des caract\'eristiques g\'en\'eralis\'ees pour les syst\`emes
hyperboliques et quasilin\'eaires du premier ordre.  Une variante de
la m\'ethode des sym\'etries conditionnelles pour obtenir ce genre
de solutions est pr\'esent\'ee.  Un module de Lie de champs de
vecteurs, repr\'esentant des symm\'etries du syst\`eme
surd\'etermin\'e constitu\'e du syst\`eme d'\'equations initial et
certaines contraintes diff\'erentielles du premier ordre, est
construit.  Il est d\'emontr\'e que ce syst\`eme surd\'etermin\'e
admet des solutions de rang k exprim\'ees en termes d'invariants de
Riemann.  Finalement, \`a titre d'exemple, une discussion
d\'etaill\'ee de l'application de l'approche propos\'ee aux
\'equations de la dynamique des fluides en (k+1) dimensions est
donn\'ee.  Plusieurs nouvelles solutions de type alg\'ebriques,
rationnelles et solitoniques (bumps, kinks et ondes multiples) ont
\'et\'e obtenues.\\[1cm]

PACS subject classification (1994) : Primary 03.40.Kf; Secondary
02.20.Sv and 02.30.Jr

\end{abstract}

\newpage
\setcounter{page}{1}
\section{Introduction}

\indent The general properties of nonlinear systems of PDEs in many
dimensions and techniques for obtaining their exact solutions remain
essential subjects of investigation in modern mathematics.  In the
case of hyperbolic systems, the oldest, and still useful, approach
to this subject has been the method of characteristics which originated
from the work of G. Monge \cite{Monge}. In its modern form it is
described e.g. in \cite{Da},\cite{J},\cite{M1},\cite{P1},\cite{RJ},\cite{Zakharov}.  More
recently, the development of group theoretical methods, based on the
work of S. Lie \cite{LE}, has led to progress in this area,
delivering new efficient techniques.  However, these two theoretical
approaches have remained disconnected and have provided, in most
cases, different sets of solutions.  The symmetry reduction methods
(SRM) certainly have a broader range of application, while the
generalized method of characteristics (GMC), though limited to
nonelliptic systems, has been more successful in producing wave and
multiple wave solutions.  Thus, the mutual relation between these
two methods is a matter of interest and we have undertaken this
subject with the view of combining the strengths of both of them.

The approach to constructing rank-k solutions presented in this paper evolved from our earlier work \cite{GMR},\cite{GT},\cite{GH}, aimed at obtaining Riemann k-waves by means of the conditional symmetry method (CSM).  The main idea here has been to select the supplementary differential constraints (DCs), employed by this method, in such a way that they ensure the existence of solutions expressible in terms of Riemann invariants.  Interestingly, as we show later, these constraints prove to be less restrictive than the conditions required by the GMC.  As a result, we obtain larger classes of solutions than the class of Riemann k-waves obtainable through the GMC.

The organization of this paper is as follows.  Section 2 gives a brief account of the generalized method of
characteristics for first order quasilinear hyperbolic systems of PDEs in many independent and dependent
variables.  A geometric formulation of the Riemann k-wave problem is presented there.  In Section 3 we reformulate this problem or rather, more generally, a problem of rank-k solutions expressible in terms of Riemann invariants, in the language of group theoretical approach.  The necessary and sufficient
conditions for obtaining this type of solutions are determined
after an analysis of their group properties.  A new version of the
conditional symmetry method for construction of these solutions is proposed.
%In Section 4, we describe the connection between this method and the method of characteristics.  
Sections 4 to 7 present an application of the developed approach to the equations describing an ideal nonstationary isentropic compressible fluid.  We find rank-1 as well as rank-2 and rank-3 solutions admitted by the system, among them several new types of soliton-like solutions including kinks, bumps and snoidal waves.  In Section 8, we construct rank-k solutions for the isentropic flow with sound velocity dependent on time only.  We show that the general integral of a Cauchy problem for this system depends on $k$ arbitrary functions of $k$ variables.  Section 9 summarizes the obtained results and contains some suggestions regarding further developments.

% In this context, we discuss some advantages of the proposed method.  Section 5 contains examples of the application of this method
% to the equations describing an ideal, nonstationary isentropic compressible fluid.  We are
% interested in the possibility of rank-1 solutions and their superpositions, i.e. rank-k solutions, admitted
% by this hydrodynamic system of equations.  To analyze this problem, we use the results obtained
% in Section 3.  It allows us to determine the types of rank-k solutions connected with the sets of
% Lie modules of certain vector fields which are generated in the process of superpositions. A
% detailed representation of relations among Lie modules is discussed in detail in Sections 6 and 7.  This results in a large number of new bounded conditionally invariant solutions.  The rank-k solutions represent algebraic solitons, kinks, bumps and snoidal waves.  In Section 8, we find the general integral of a Cauchy problem for the isentropic flow with time dependent sound velocity and show that it depends on $n$ arbitrary functions of $n$ variables. Section 9 summarizes the obtained results and contains suggestions regarding possible
% developments.

\section{The generalized method of characteristics}

\indent The generalized method of characteristics has been designed for the purpose of solving 
quasilinear hyperbolic systems of first order PDEs in many dimensions.This approach enables us to 
construct and investigate Riemann waves and their superpositions (i.e. Riemann k-waves), which are 
admitted by these systems.  The main feature of the method is the introduction of new independent variables 
(called Riemann invariants) which remain constant on certain hyperplanes perpendicular to wave vectors associated 
with the initial system.  This results in a reduction of the dimensionality of the problem. A number of attempts to 
generalize the Riemann invariants method and its various applications can be found in the recent literature of the 
subject (see e.g.  \cite{D1} - \cite{F2}, \cite{Mokhov}, \cite{Pa} and references therein).

At this point, we summarize the version of the GMC for constructing k-wave solutions developed progressively in  \cite{B2}, \cite{GZ}, \cite{GV}, \cite{P4}, \cite{P1}.
Let us consider a quasilinear hyperbolic system of $l$ first order PDEs
\begin{equation}
 \label{quasi}
{\CA^{\mu}}^i_{\alpha} (u) u^{\alpha}_i = 0, \quad \mu=1,\ldots,l,\quad \alpha=1,\ldots,q, \quad i=1,\ldots,p,
\end{equation}
in $p$ independent variables $x = (x^1, \ldots, x^p) \in X \subset \mathbb{R}^p$ and $q$ dependent variables $u = (u^1, \ldots, u^q) \in U \subset \mathbb{R}^q$.  The term $u^{\alpha}_i$ denotes the first order partial derivative of $u^{\alpha}$ with respect to $x^i$, i.e. $u^{\alpha}_i \equiv  \p u^{\alpha} / \p x^i$.  Here we adopt the summation convention over the repeated lower and upper indices, except in the cases in which one index is taken in brackets.  The system is properly determined if $l=q$.  All considerations have local character, that is, it suffices to look for solutions defined in a neighborhood of $x=0$.  The main steps in constructing k-wave solutions can be presented as follows.

\noindent {\bf 1.} Find the real-valued functions $\lambda^A = (\lambda^A_1,\ldots,\lambda^A_p)
\in X$ and $\gamma_A = (\gamma_A^1,\ldots,\gamma_A^q) \in U$ by solving the wave relation
associated with the initial system (\ref{quasi}),
\begin{equation}
\label{GMC-wave-relation}
(\CA^{\mu}{}^i_{\alpha}(u) \lambda^A_i)\gamma^{\alpha}_{(A)} = 0, \quad  A = 1,\ldots, k < p.
\end{equation}
Thus we  require that the condition
\begin{equation}
\label{GMC-rank-cond}
\rank{(\CA^{\mu}{}^i_{\alpha}(u)\lambda^A_i)}<\min{(l,q)}
\end{equation}
holds.  We assume here the generic case in which the rank does not vary on some open
subset $\Omega \subset U$.  This step is completely algebraic.\\
{\bf 2.} Let us assume that we have found $k$ linearly independent functions $\lambda^A$ and
$\gamma_A$ which are $C^1$ in $\Omega$.  We postulate a form of solution $u(x)$ of the initial system (\ref{quasi})
such that all first order derivatives of $u$ with respect to $x^i$ are decomposable in the following way
\begin{equation}
\label{GMC-DC}
\frac{\p u^{\alpha}}{\p x^i}(x) = \sum_{A=1}^k h^A(x) \gamma^{\alpha}_{A}(u) \lambda^A_i(u)
\end{equation}
on some open domain ${\cal B} \subset X \times U$.  Here, $h^A(x)$
are arbitrary functions of $x$.  This step means that the
original system (\ref{quasi}) is subjected to the first order
differential constraints (\ref{GMC-DC}).  Thus we have to solve an
overdetermined system composed of (\ref{quasi}) and (\ref{GMC-DC}).

The condition (\ref{GMC-DC}), crucial to the GMC, determines the
class of solutions, called Riemann k-waves, resulting from superposition of k simple waves.

\noindent {\bf 3.}  Before proceeding further, we should verify whether the conditions on the vector functions $\lambda^A$ and $\gamma_A$, which are necessary and sufficient for the existence of solutions of the system composed of (\ref{quasi}) and (\ref{GMC-DC}), are satisfied.  These conditions, in accordance with the Cartan theory of systems in involution \cite{C}, take the form
\begin{equation}
\label{GMC-existence}
\mathrm{i)}\, [\gamma_A,\gamma_B] \in \mathrm{span} \{\gamma_A,\gamma_B\}, \quad \mathrm{ii)}\, {\cal L}_{\gamma_B} \lambda^A \in \mathrm{span}\{\lambda^A, \lambda^B\}, \quad A\neq B
= 1, \ldots, k,
\end{equation}
where ${\cal L}_{\gamma_B}$ denotes the Lie derivative along the vector field $\gamma_B$ and the
bracket $[\gamma_A,\gamma_B]$ denotes the commutator of the vector fields $\gamma_A, \gamma_B$.\\
{\bf 4.} Given that the conditions (\ref{GMC-existence}) are satisfied, we can choose, due to the homogeneity of the wave relation (\ref{GMC-wave-relation}), a holonomic system for the vector fields $\{\gamma_1,\ldots,\gamma_k\}$, by requiring a proper length
for each vector $\gamma_A$ such that
\begin{equation}
\label{GMC-commut}
[\gamma_A,\gamma_B] = 0.
\end{equation}
Conditions (\ref{GMC-commut}) determine a k-dimensional submanifold ${\cal S} \subset U$ which can be obtained by solving the system of
PDEs
\begin{equation}
\label{GMC-sub-PDEs}
\frac{\p u^{\alpha}}{\p r^A} = \gamma^{\alpha}_A (u^1,\ldots, u^q), \quad A=1,\ldots, k.
\end{equation}
The solution of (\ref{GMC-sub-PDEs})
\begin{equation}
\label{GMC-sub-PDEs-sol}
u = f(r^1,\ldots,r^k)
\end{equation}
gives the explicit parametrization of the submanifold ${\cal S}$ immersed in the space $U$ .\\
{\bf 5.} Next we consider the functions $f^*(\lambda^A)$, that is, the functions $\lambda^A(u)$
pulled back to the submanifold ${\cal S} \subset U$.  The $\lambda^A(u)$ become functions of the
parameters $(r^1,\ldots, r^k)$ on ${\cal S}$.  For simplicity of notation, we denote
$f^*(\lambda^A)$ by $\lambda^A(r^1,\ldots, r^k)$.

\noindent {\bf 6.} Restricting the equations (\ref{GMC-DC}) and (\ref{GMC-existence}) to the
submanifold ${\cal S}$ and using the linear independence of the vectors $\{\gamma_1, \ldots,
\gamma_k \}$, we obtain
\begin{eqnarray}
\label{GMC-pull-req-a}
& &\frac{\p r^A}{\p x^i} = h^A(x) \lambda^A_i(r^1,\ldots,r^k),\\
\label{GMC-pull-req-b}
& &\frac{\p \lambda^A}{\p r^B} = \alpha^A_B(r^1,\ldots, r^k) \lambda^A + \beta^A_B(r^1,\ldots,r^k) \lambda^B, \quad A \neq B = 1,\ldots,k,
\end{eqnarray}
for some real-valued functions $\alpha^A_B$ and $\beta^A_B$ $ : {\cal S} \to \NR$.  Here we do not use the summation convention.
According to the Cartan theory of systems in involution, the conditions
(\ref{GMC-sub-PDEs}), (\ref{GMC-pull-req-a}) and (\ref{GMC-pull-req-b}) ensure that the set
of solutions of the initial system of PDEs (\ref{quasi}) subjected to the differential constraints (\ref{GMC-DC}) depends on $k$ arbitrary functions of one
variable.\\
{\bf 7.} Next, we look for the most general class of solutions of the linear system of equations
(\ref{GMC-pull-req-b}) for $\lambda^A$ as functions of $r^1,\ldots, r^k$.  We can perform this
analysis by using, for example, the Monge-Darboux method \cite{GV}.\\
{\bf 8.} From the general solution of (\ref{GMC-pull-req-b}) for the functions $\lambda^A$, the
solution of the system (\ref{GMC-pull-req-a}) can be derived in the implicit form
\begin{equation}
\label{GMC-lambda-psi}
\lambda^A_i (r^1,\ldots, r^k) x^i = \psi^A (r^1,\ldots, r^k), \quad A=1, \ldots, k,
\end{equation}
where $\psi^A : \NR^k \to \NR$ are some functionally independent differentiable functions of
$k$ variables $r^1,\ldots, r^k$ such that
\begin{equation}
\label{psi_rb}
\frac{\p \psi^A}{\p r^B} = \alpha^A_B(r^1, \ldots, r^k) \psi^{(A)} + \beta^A_B(r^1, \ldots, r^k) \psi^{(B)}, \quad A\neq B.
\end{equation}
Note that the solutions $r^A(x)$ of (\ref{GMC-lambda-psi}) are constant
on (p-k)-dimensional hyperplanes perpendicular to the wave vectors $\lambda^A$.\\
{\bf 9.} Finally, the k-wave solution of (\ref{quasi}) is obtained from the explicit parametrization (\ref{GMC-sub-PDEs-sol}) of the submanifold ${\cal S} \subset U$ in terms of the parameters $r^1,\ldots, r^k$, which are now implicitly defined as functions of $x^1,\ldots,x^p$ by the solutions
of the system (\ref{GMC-lambda-psi}) in the space $X$ .

If the set of implicitly defined relations between the variables $u^{\alpha}$,
$x^i$ and $(r^1,\ldots,r^k)$,
\begin{equation}
u^{\alpha} = f^{\alpha} (r^1,\ldots,r^k), \quad \lambda^A_i (r^1,\ldots,r^k)x^i =
\psi^A(r^1,\ldots, r^k),
\end{equation}
can be solved in such a way that $r$ and $u^{\alpha}$ can be given as graphs over an open subset ${\cal D}
\subset X$,  then the functions $u^{\alpha} = f^{\alpha}(r^1(x),\ldots,r^k(x))$ constitute a k-wave
solution of the quasilinear hyperbolic system (\ref{quasi}).  The scalar functions $r^A(x)$ are
called the Riemann invariants.  For $p=2$ they coincide with the classical Riemann invariants as they have 
been usually introduced in the literature of the subject (see e.g. \cite{CH},\cite{J},\cite{RJ},\cite{W1}).

Finally, let us comment on the Cauchy problem for Riemann k-waves (for a detailed discussion see e.g. \cite{GV},\cite{M1},\cite{P1}).

Let us consider $q$ functions $u^1(x),\ldots,u^q(x)$ which take some prescribed values $u_0(\bar{x}) = (u^1_0(\bar{x}),\ldots,u^q_0(\bar{x}))$ on the hyperplane $\mathbb{P} \subset \NR^n$ defined by $t=0$.  Here, we use the notation $x=(t,\bar{x}) \in X \subset \NR^{n+1}$.  It was shown \cite{GV} that for $0 < t <T$ the initial value problem for the system (\ref{quasi}) has locally exactly one solution in the form of a Riemann k-wave defined implicitly by relations (\ref{GMC-sub-PDEs-sol}), (\ref{GMC-lambda-psi}) and (\ref{psi_rb}) if the $C^2$ function $u_0(\bar{x})$ satisfies the following conditions.\\
{\bf i) } $u_0(\bar{x})$ is sufficiently small that there exists a time interval $[0,T]$ in which the gradient catastrophe for a solution $u(x)$ of (\ref{quasi}) does not occur.\\
{\bf ii) } $u_0(\bar{x})$ is decomposable according to conditions (\ref{GMC-DC}), that is
\begin{equation}
\frac{\p u^{\alpha}}{\p x^j} (\bar{x}) \Big|_{\mathbb{P}} = \sum_{A=1}^k \xi^A(0,\bar{x}) \gamma^{\alpha}_A(u_0(\bar{x})) \lambda^A_j(u_0(\bar{x})) \Big|_{\mathbb{P}}
\end{equation}
on some open domain $\mathcal{E} \subset \mathbb{P} \times U$.

\section{Conditional Symmetries and Riemann Invariants}

Until now, the only way to approach the problem of superposition of many Riemann waves in multi-dimensional space was through the GMC.  
This method, like all other techniques of solving PDEs, has its limitations.  They have motivated the authors to search for the means of 
constructing larger classes of multiple wave solutions expressible in terms of Riemann invariants.  The natural way to do it is to look at 
these solutions from the point of view of group invariance properties.  The feasibility and advantages of such an approach were demonstrated 
for certain fluid dynamic equations in \cite{GT},\cite{GH}.  We have been particularly interested in the construction of nonlinear superpositions of 
elementary solutions (i.e. rank-1 solutions) of (\ref{quasi}), and the preliminary analysis indicated that the method of conditional symmetry 
is an especially useful tool for this purpose.

We use the term ``conditional symmetry'' here as introduced by P.J. Olver and P. Rosenau \cite{OR}.  It evolved from the notion of 
``nonclassical symmetry'' which had originated from the work of G. Bluman and J. Cole \cite{BC} and was developed by several authors 
(D. Levi and P. Winternitz \cite{LW} and Fushchych \cite{Fu} among others).  For a review of this subject see e.g. \cite{BK}, \cite{CW}, \cite{OV} and references therein.

The method of conditional symmetry consists in supplementing the original system of PDEs with first order differential constraints 
for which a symmetry criterion of the given system of PDEs is identically satisfied.  Under certain circumstances this augmented system of 
PDEs admits a larger class of Lie symmetries than the original system of PDEs.  For our purpose we adapt here the version of CSM developed in \cite{GMR}, \cite{GT1}.

We now reformulate the task of constructing rank-k solutions expressible in terms of Riemann invariants in the language of the group theoretical approach.  Let us consider the nondegenerate system (\ref{quasi}) in its matrix form
\begin{equation}
\label{quasimatrix}
\CA^1(u) u_1 + \ldots + \CA^p(u) u_p = 0,
\end{equation}
where $\CA^1, \ldots, \CA^p$ are $l$ by $q$ real-valued matrix functions of $u$.  If we set $l=q$, $p=n+1$ (we
denote the independent variables by $t=x^0, x^1,\ldots, x^n$) and $\CA^0$ is the identity matrix,
then the system has the evolutionary form
\begin{equation}
\label{quasievo}
u_t + \sum_{j=1}^n \CA^j(u) u_j = 0.
\end{equation}
For a fixed set of $k$ linearly independent real-valued wave vectors 
\begin{equation*}
\lambda^A(u) = (\lambda^A_1(u), \ldots, \lambda^A_p(u)), \quad A=1,\ldots,k < p,
\end{equation*}
with
\begin{equation}
\label{xivec2}
\ker{(\lambda^A_i \CA^i)} \neq 0,
\end{equation}
we define the real-valued functions $r^A : X \times U \to \NR$ such that
\begin{equation}
\label{rank-k-riemann-invariants}
 r^A(x,u) = \lambda^A_i(u) x^i, \quad A=1, \ldots, k.
\end{equation}
These functions are Riemann invariants associated with the wave vectors $\lambda^A$, as introduced in the previous section.

We postulate the form of solution of (\ref{quasimatrix}) defined implicitly by the following set of relations between the variables $u^{\alpha}$, $x^i$ and $r^A$ 
\begin{equation}
\label{implicit-equations-k}
u = f(r^1(x,u), \ldots, r^k(x,u)),\quad
r^A(x,u) = \lambda^A_i(u) x^i, \quad A=1, \ldots, k.
\end{equation}
Equations (\ref{implicit-equations-k}) determine a unique function $u(x)$ on a neighborhood of $x=0$ for any $f : \NR^k \to \NR^q$.  The Jacobi matrix of equations (\ref{implicit-equations-k}) can be presented as
\begin{equation}
\label{Jacobi}
\p u = (u^{\alpha}_i) = \left({\cal I}_q - \frac{\p f}{\p r} \cdot \frac{\p r}{\p u}\right)^{-1} \frac{\p f}{\p r}
\lambda \, \in \NR^{q \times p},
\end{equation}
or equivalently as
\begin{equation}
\label{Jacobi2}
  \p u = \frac{\p f}{\p r}\left({\cal I}_k - \frac{\p r}{\p u} \cdot
  \frac{\p f}{\p r}\right)^{-1} \lambda \, \in \NR^{q \times p},
\end{equation}
where 
\begin{eqnarray}
\label{threematrices}
& &\frac{\p f}{\p r} = \left(
\frac{\p f^{\alpha}}{\p r^A}\right) \in \NR^{q \times k}, \quad \lambda = (\lambda^A_i) \in
\NR^{k \times p},\\
\label{drdu}
& &\frac{\p r}{\p u} = \left(\frac{\p r^A}{\p u^{\alpha}}\right) = \left(\frac{\p \lambda^A_i}{\p
 u^{\alpha}} x^i\right) \in \NR^{k \times q},\quad r = (r^1,\ldots, r^k) \in
 \NR^k,
\end{eqnarray}
and ${\cal I}_q$ and ${\cal I}_k$ are the $q$ by $q$ and $k$ by
$k$ identity matrices respectively.
Applying the implicit function theorem, we obtain the following conditions ensuring that $r^A$ and $u^{\alpha}$ are expressible as graphs over some open subset ${\cal D}$ of $\NR^p$,
\begin{equation}
\label{implicitcondition} \det{\left({\cal I}_q - \frac{\p f}{\p r}
\cdot \frac{\p \lambda}{\p u} x\right)} \neq 0,
\end{equation}
or
\begin{equation}
\label{implicitcondition2} \det{\left({\cal I}_k - \frac{\p
\lambda}{\p u} x  \cdot \frac{\p f}{\p r}\right)} \neq 0.
\end{equation}
The inverse matrix in (\ref{Jacobi}) (or in (\ref{Jacobi2})) is well defined, since
\begin{equation}
\frac{\p r}{\p u} = 0 \quad \text{at} \quad x=0.
\end{equation}

In our further considerations we assume that the conditions (\ref{implicitcondition}) or (\ref{implicitcondition2}) are fulfilled, whenever applicable.

The postulated solution (\ref{implicit-equations-k}) is a rank-k solution, since the Jacobi matrix of $u(x)$ has a rank equal to $k$.  Its image is a k-dimensional submanifold ${\cal S}_k$ in the first jet space $J^1 = J^1(X \times U)$.

For a fixed set of $k$ linearly independent wave vectors $\{\lambda^1, \ldots, \lambda^k\}$ we define another set of $(p-k)$ linearly independent vectors
\begin{equation}
 \xi_a (u) = (\xi^1_a(u), \ldots, \xi^p_a(u))^T, \quad a=1,\ldots,p-k,
\end{equation}
satisfying the orthogonality conditions
\begin{equation}
\label{xivec}
 \lambda^A_i \xi^i_a = 0, \quad A=1,\ldots,k.
\end{equation}
Then, due to (\ref{Jacobi}) (or (\ref{Jacobi2})), the graph of the solution $\Gamma = \{(x,u(x))\}$ is invariant under the family of the first order differential operators
\begin{equation}
\label{diffop}
X_a = \xi^i_a(u) \frac{\p}{\p x^i}, \quad a=1,\ldots,p-k,
\end{equation}
defined on $X \times U$.  Note that the vector fields $X_a$ do not include vectors tangent to the direction $u$.  So, the vectors fields $X_a$ form an Abelian distribution on $X \times U$, i.e.
\begin{equation}
\label{V-F-commute}
[X_a , X_b] = 0, \quad a \neq b = 1,\ldots, p-k.
\end{equation}

Conversely, if $u(x)$ is a q-component function defined on a neighborhood of $x=0$ such
that the graph $\Gamma = \{(x,u(x))\}$ is invariant under a set of $(p-k)$ vector fields $X_a$
with properties (\ref{xivec}), then $u(x)$ is a solution of equations (\ref{implicit-equations-k}), for some $f$.  This is so, because the set $\{r^1,\ldots,r^k, u^1,\ldots, u^q\}$ constitutes a complete
set of invariants of the Abelian algebra of the vector fields (\ref{diffop}).  This geometrically
characterizes the solutions $u(x)$ of the equations (\ref{implicit-equations-k}).

The group-invariant solutions of the system
(\ref{quasimatrix}) consist of those functions $u=f(x)$ which satisfy both the initial system
(\ref{quasimatrix}) and a set of first order differential constraints
\begin{equation}
\label{rank-k-DCs}
 \xi^i_a u^{\alpha}_i = 0, \quad a=1,\ldots, p-k, \quad \alpha=1,\ldots,q,
\end{equation}
ensuring that the characteristics of the vectors fields $X_a$ are equal to zero.  

Note that, in general, the conditions (\ref{rank-k-DCs}) are weaker than the DCs (\ref{GMC-DC}) required by the GMC, since the latter are submitted to the algebraic condition (\ref{GMC-wave-relation}).  Indeed, (\ref{rank-k-DCs}) implies
\begin{equation}
\label{jet-3}
u^{\alpha}_i = \Phi^{\alpha}_A \lambda^A_i,
\end{equation}
where $\Phi^{\alpha}_A$ are real-valued matrix functions on the first jet space $J^1 = J^1(X \times U)$,
\begin{equation}
\label{zalpha-a}
\Phi^{\alpha}_A = \left[\left({\cal I}_q - \frac{\p f}{\p r} \cdot \frac{\p r}{\p u}\right)^{-1}
\right]^{\alpha}_{\beta} \frac{\p f^{\beta}}{\p r^A} \in \NR^{q \times k},
\end{equation}
or
\begin{equation}
\label{zalpha-b}
\Phi^{\alpha}_A = \frac{\p f^{\alpha}}{\p r^B} \left[\left({\cal I}_k - \frac{\p r}{\p u}
\cdot \frac{\p f}{\p r}\right)^{-1}\right]^B_A \in \NR^{q \times k},
\end{equation}
which do not necessarily satisfy the wave relation (\ref{GMC-wave-relation}).  This fact results in easing up the restrictions on initial data at $t=0$, thus we are able to consider more diverse configurations of waves involved in a superposition than in the GMC case.

We now proceed to solve the overdetermined system composed of the initial system (\ref{quasimatrix}) and the DCs (\ref{rank-k-DCs})
\begin{equation}
\label{over-system-DC} \CA^i{}^{\mu}_{\alpha}(u) u^{\alpha}_i = 0,
\quad \xi^i_a (u) u^{\alpha}_i = 0, \quad \mu = 1, \ldots, l, \quad
a=1,\ldots, p-k.
\end{equation}
Substituting (\ref{Jacobi}) (or (\ref{Jacobi2})) into (\ref{quasimatrix})
yields
\begin{equation}
\label{traceeq}
\tr\left(\CA^{\mu} \left({\cal I}_q - \frac{\p f}{\p r}\cdot\frac{\p r}{\p u}\right)^{-1}
\frac{\p f}{\p r} \lambda\right) = 0, \quad \mu = 1,\ldots, l,
\end{equation}
or
\begin{equation}
\label{traceeq-b} \tr\left(\CA^{\mu} \frac{\p f}{\p r} \left({\cal
I}_k - \frac{\p r}{\p u}\cdot\frac{\p f}{\p r}\right)^{-1} \lambda\right)
= 0, \quad \mu = 1,\ldots, l,
\end{equation}
where $\CA^1,\ldots, \CA^l$ are $p$ by $q$ matrix functions of $u$ (i.e. $\CA^{\mu} = \left({\CA^{\mu}}^i_{\alpha}(u)\right) \in \NR^{p \times q}$, $\mu=1,\ldots,l$).
For the given system of equations (\ref{quasi}), the matrices
$\CA^{\mu}$ are known functions of $u$ and equations (\ref{traceeq})
(or (\ref{traceeq-b})) constitute conditions on functions $f^{\alpha}(r)$ and $\lambda^A(u)$
(or, by virtue of (\ref{xivec}), on $\xi_a(u)$). It is convenient
from a computational point of view to split $x^i$ into $x^{i_A}$
and $x^{i_a}$ and to choose a basis for the wave vectors $\lambda^A$
such that
\begin{equation}
\label{lambdabasis}
\lambda^A = dx^{i_A} + \lambda^A_{i_a} dx^{i_a}, \quad A=1,\ldots,k,
\end{equation}
where $(i_A,i_a)$ is a permutation of $(1,\ldots,p)$.  Hence, expression (\ref{drdu}) becomes
\begin{equation}
\label{drdu2}
\frac{\p r^A}{\p u^{\alpha}} = \frac{\p \lambda^A_{i_a}}{\p u^{\alpha}} x^{i_a}.
\end{equation}
Substituting (\ref{drdu2}) into (\ref{traceeq}) (or (\ref{traceeq-b}))
yields
\begin{equation}
\label{traceeq2}
\tr\left(\CA^{\mu}\left({\cal I}_q - Q_a x^{i_a}\right)^{-1}\frac{\p f}{\p r} \lambda\right) = 0,
\quad \mu=1,\ldots, l,
\end{equation}
or
\begin{equation}
\label{traceeq2-b} \tr\left(\CA^{\mu} \frac{\p f}{\p r} \left({\cal
I}_k - K_a x^{i_a}\right)^{-1} \lambda\right) = 0, \quad
\mu=1,\ldots, l,
\end{equation}
where
\begin{equation}
\label{QaKa}
Q_a = \frac{\p f}{\p r} \eta_a \in \NR^{q \times
q},\quad K_a = \eta_a \frac{\p f}{\p r} \in \NR^{k \times k}, \quad
\eta_a = \left( \frac{\p \lambda^A_{i_a}}{\p u^{\alpha}}\right) \in
\NR^{k \times q},
\end{equation}
for $i_A$ fixed and $i_a = 1, \ldots, p-1$. Note that the functions
$r^A$ and $x^{i_a}$ are functionally independent in a neighborhood
of $x=0$.  The matrix functions $\CA^{\mu}$, $\p f / \p r$,
$Q_a$ and $K_a$ depend on $r$ only.  Hence, equations (\ref{traceeq2})
(or (\ref{traceeq2-b})) have to be satisfied for any value of coordinates $x^{i_a}$.  This
requirement leads to some constraints on these matrix functions.

According to the Cayley-Hamilton theorem, for any $n$ by $n$ invertible matrix $M$,
the expression
$(M^{-1} \det{M})$ is a polynomial in $M$ of order $(n-1)$.  Thus, using the tracelessness of the expression $\CA^{\mu} \left({\cal I}_q - Q_a x^{i_a}\right)^{-1} (\p f / \p r) \lambda$, we can replace equations
(\ref{traceeq2})  by the following
\begin{equation}
\label{traceeq3} \tr\left(\CA^{\mu} Q \frac{\p f}{\p r} \lambda\right)
= 0, \quad\text{where} \quad Q = \mathrm{adj} ({\cal I}_q - Q_a x^{i_a}) \in \NR^{q \times q}.
\end{equation}
Here $\mathrm{\,adj} M$ denotes the classical adjoint of the matrix
$M$.  Note that $Q$ is a polynomial of order $(q-1)$ in $x^{i_a}$.
Taking (\ref{traceeq3}) and all its partial derivatives with respect
to $x^{i_a}$ (with $r$ fixed at $x=0$), we obtain the
following conditions for the matrix functions $f(r)$ and
$\lambda(f)$
\begin{eqnarray}
\label{traceeq4}
&&\tr\left(\CA^{\mu}\frac{\p f}{\p r} \lambda\right) = 0,\quad \mu=1,\ldots,l,\\
\label{traceeq5-a}
&&\tr\left(\CA^{\mu} Q_{(a_1} \ldots Q_{a_s)}\frac{\p f}{\p r} \lambda\right) = 0,
\end{eqnarray}
where $s=1,\ldots,q-1$ and $(a_1,\ldots,a_s)$ denotes the
symmetrization over all indices in the bracket.  A similar procedure
can be applied to system (\ref{traceeq2-b}) to yield
(\ref{traceeq4}) and
\begin{equation}
\label{traceeq5-b}
  \tr\left(\CA^{\mu} \frac{\p f}{\p r} K_{(a_1} \ldots K_{a_s)} \lambda\right) = 0, \quad K=\mathrm{adj} ({\cal I}_k - K_a x^{i_a}) \in \NR^{k\times k},
\end{equation}
where now $s=1,\ldots,k-1$.
Equations (\ref{traceeq4}) represent the initial value conditions on a
surface in the space of independent variables $X$, given at
$x^{i_a}=0$.  Note that equations (\ref{traceeq5-a}) (or
(\ref{traceeq5-b})) form the conditions required for preservation of the property
(\ref{traceeq4}) by flows represented by the vector fields
(\ref{diffop}).  Note also that, by virtue of (\ref{lambdabasis}), $X_a$ can be expressed in the form
\begin{equation}
  X_a = \p_{i_a} - \lambda^A_{i_a} \p_{i_A}.
\end{equation}
Substituting expressions (\ref{QaKa}) into (\ref{traceeq5-a}) or (\ref{traceeq5-b}) and simplifying gives the unified form
\begin{equation}
\label{traceeq5}
\tr\left(\CA^{\mu}\frac{\p f}{\p r} \eta_{(a_1} \frac{\p f}{\p r} \ldots \eta_{a_s)}
\frac{\p f}{\p r} \lambda\right) = 0, \quad \eta_{a_t} = \left(\frac{\p \lambda^A_{a_t}}{\p u^{\alpha}} \right)
\in \NR^{k \times q}, \quad t=1,\ldots,s,
\end{equation}
where we can choose either $\max{(s)} = q-1$ or $\max{(s)}=k-1$, whichever is more convenient.

Let us note that for $k=1$ the results of the two methods, CSM and GMC, overlap.  This is due to the fact that conditions (\ref{GMC-wave-relation}) and (\ref{GMC-sub-PDEs}) coincide with (\ref{traceeq4}) and conditions (\ref{GMC-existence}) and (\ref{traceeq5}) are identically equal to zero.  In this case, all rank-1 solutions correspond to single Riemann waves.  However, for $k\geq 2$ the differences between the two approaches become essential and, as we demonstrate in the following examples, the CSM can provide rank-k solutions which are not Riemann k-waves as defined by the GMC.

We now introduce a change of variables on $\NR^p \times \NR^q$ which allows us to rectify the vector fields $X_a$ and simplify considerably the structure of the overdetermined system (\ref{over-system-DC}).  For this system, in the new coordinates, we derive the necessary and sufficient conditions for existence of rank-k solutions in the form (\ref{implicit-equations-k}).

Let us assume that there exists an invertible $k$ by $k$ subblock
\begin{equation}
\label{k-k-invertible}
\Lambda = (\lambda^A_B), \quad 1 \leq A,B \leq k,
\end{equation}
of the matrix $\lambda \in \NR^{k \times p}$.  Then the independent vector fields
\begin{equation}
\label{rank-k-vf} X_{k+1} = \frac{\p}{\p x^{k+1}} -
(\Lambda^{-1})^B_A \lambda^A_{k+1} \frac{\p}{\p x^B},\quad \dots
\quad , X_p = \frac{\p}{\p x^p} -  (\Lambda^{-1})^B_A \lambda^A_p
\frac{\p}{\p x^B},
\end{equation}
have the required form (\ref{diffop}) for which the orthogonality conditions (\ref{xivec}) are satisfied.  We introduce the functions
\begin{equation}
\label{rank-k-coord}
\begin{split}
&\bx^1 = r^1(x,u),\, \ldots,\, \bx^k = r^k(x,u), \\
&\bx^{k+1} = x^{k+1},\, \ldots,\, \bx^p = x^p, \bu^1 = u^1,\, \ldots,\, \bu^q = u^q, \end{split}
\end{equation}
as new coordinates on $\NR^p \times \NR^q$ space which allow us to rectify the vector fields
(\ref{rank-k-vf}).  So, we get
\begin{equation}
\label{rank-k-vf-rect}
X_{k+1} = \frac{\p}{\p \bx^{k+1}}, \ldots, X_p = \frac{\p}{\p \bx^p}.
\end{equation}
The p-dimensional submanifold invariant under $X_{k+1},\ldots, X_p$ is defined by
equations of the form
\begin{equation}
\label{rank-k-vf-sol}
\bu = f(\bx^1,\ldots,\bx^k),
\end{equation}
for an arbitrary function $f : \NR^k \to \NR^q$.  The expression (\ref{rank-k-vf-sol}) is the general
solution of the invariance conditions
\begin{equation}
\label{rank-k-inv}
\bu_{\bx^{k+1}}, \ldots, \bu_{\bx^p} = 0.
\end{equation}
The initial system (\ref{quasimatrix}) described in the new coordinates $(\bx,\bu) \in \NR^p \times
\NR^q$ is, in general, a nonlinear system of first order PDEs,
\begin{equation}
\label{rank-k-syst-trans}
\bar{\CA}^i(\bx,\bu,\bu_{\bx}) \bu_i = 0, \quad \text{where} \quad
\bar{\CA}^i = \frac{\p \bx^i}{\p x^j} \CA^j,\quad i,j = 1, \ldots, p.
\end{equation}
That is, we have
\begin{equation}
\label{rank-k-matrix}
\bar{\CA}^1 = \frac{\p r^1}{\p x^i} \CA^i, \ldots, \bar{\CA}^k = \frac{\p r^k}{\p x^i} \CA^i,
\bar{\CA}^{k+1} = \CA^{k+1}, \ldots, \bar{\CA}^p = \CA^p.
\end{equation}
The Jacobi matrix in the coordinates $(\bx,\bu)$ takes the form
\begin{equation}
\label{rank-k-jacob-trans}
\frac{\p r^i}{\p x^j} = (\phi^{-1})^i_s \lambda^s_j \in \NR^{k \times p}, \quad
(\phi^i_j) = \left(\delta^i_j - \frac{\p r^i}{\p u^l} \frac{\p \bu^l}{\p \bx^j}
\right) \in \NR^{k \times k},
\end{equation}
whenever the invariance conditions (\ref{rank-k-inv}) are satisfied.  Augmenting the system
(\ref{rank-k-syst-trans}) with the invariance conditions (\ref{rank-k-inv}) leads to a quasilinear
reduced system of PDEs
\begin{equation}
\label{rank-k-syst-trans-inv} \Delta : \begin{cases} &
{\displaystyle \tr{\left({\bar{\CA}}^{\mu} \left( {\cal I}_q - \frac{\p
\bu}{\p \bx} \cdot \frac{\p r}{\p u} \right)^{-1} \frac{\p
\bu}{\p \bx} \lambda\right)} = 0, \quad
\mu = 1, \ldots, l,}\\
& \bu_{\bx^{k+1}}, \ldots, \bu_{\bx^p} = 0,
\end{cases}
\end{equation}
or
\begin{equation}
\label{rank-k-syst-trans-inv-b} \Delta : \begin{cases} &
{\displaystyle \tr{\left({\bar{\CA}}^{\mu}  \frac{\p
\bu}{\p \bx} \left( {\cal I}_k - \frac{\p r}{\p u} \cdot \frac{\p
\bu}{\p \bx} \right)^{-1}  \lambda\right)} = 0, \quad
\mu = 1, \ldots, l,}\\
& \bu_{\bx^{k+1}}, \ldots, \bu_{\bx^p} = 0.
\end{cases}
\end{equation}

Now we present some examples which illustrate the preceeding construction.  If
\begin{equation}
\phi = {\cal I}_q - \frac{\p \bu}{\p \bx}\cdot \frac{\p r}{\p u} \in \NR^{q \times q}
\end{equation}
is a scalar matrix, then system (\ref{rank-k-syst-trans-inv}) is equivalent to the following
quasilinear system
\begin{equation}
\label{Comp-quasi}
B^i(\bu)\bu_i = 0
\end{equation}
in $k$ independent variables $\bx^1,\ldots,\bx^k$ and $q$ dependent variables $\bu^1,\ldots,\bu^q$,
where $B^i = \lambda^i_j \CA^j$.  In the simplest case, when $k=1$, the equations
(\ref{Comp-quasi}) coincide with the system (\ref{rank-k-syst-trans-inv}), i.e.
\begin{equation}
\lambda_i (\bu) \CA^i(\bu) \bu_1 = 0, \quad \bu^2, \ldots, \bu^p = 0,
\end{equation}
with the general solution
\begin{equation}
\bu(\bx) = f(\bx^1),
\end{equation}
where $f : \NR \to \NR^q$ satisfies the first order ordinary differential equation 
\begin{equation}
\lambda_i(f) \CA^i(f) f' = 0,
\end{equation}
and we have used the following notation $f' = df/d\bx^1$.

If $k\geq 2$ then $\phi$ is a scalar matrix if and only if the Riemann invariants do not depend on the function $u$,
\begin{equation}
\label{Comp-nodepends-u}
\frac{\p r^1}{\p u}, \ldots, \frac{\p r^k}{\p u} = 0.
\end{equation}
Consequently, equations (\ref{drdu2}) and (\ref{Comp-nodepends-u}) imply that the wave vectors
$\lambda^1, \ldots, \lambda^k$ are constant.  Hence, this solution represents a travelling k-wave.

Consider now a more general situation when the matrix $\phi$ does not depend on variables
$\bx^{k+1}, \ldots, \bx^p$, that is
\begin{equation}
\frac{\p \phi}{\p \bx^{k+1}}, \ldots, \frac{\p \phi}{\p \bx^p} = 0.
\end{equation}
The system (\ref{rank-k-syst-trans-inv}) is independent of $\bx^{k+1}, \ldots, \bx^p$ if and only if
\begin{equation}
\frac{\p^2 r}{\p u \p \bx^{k+1}}, \ldots, \frac{\p^2 r}{\p u \p \bx^p} = 0,
\end{equation}
or equivalently, due to (\ref{k-k-invertible}), if and only if
\begin{equation}
\frac{\p \lambda^A_i}{\p u} = \frac{\p \Lambda^A_m}{\p u} (\Lambda^{-1})^m_n
\lambda^n_i, \quad 1 \leq A \leq k < i < p, \quad m,n = 1,\ldots,k.
\end{equation}
So, it follows that
\begin{equation}
\frac{\p}{\p u} \left( (\Lambda^{-1})^A_m \lambda^m_i\right) = 0, \quad 1 \leq A
\leq k < i < p.
\end{equation}
Thus, equations (\ref{rank-k-syst-trans-inv}) are independent of $\bx^{k+1}, \ldots, \bx^p$ if
there exists a $k$ by $(p-k)$ constant matrix $C$ such that
\begin{equation}
(\lambda^A_i) = \Lambda \cdot C, \quad 1 \leq A \leq k < i < p.
\end{equation}
In this case, (\ref{rank-k-syst-trans-inv}) is a system (not necessarily a quasilinear one) in $k$
independent variables $\bx^1, \ldots, \bx^k$ and $q$ dependent variables $\bu^1,\ldots, \bu^q$.

Let us now proceed to define some basic notions of the conditional
symmetry method in the context of Riemann invariants.  We associate
the original system (\ref{quasimatrix}) and the invariance
conditions (\ref{rank-k-DCs}) with the subvarieties of the solution
spaces
\begin{eqnarray*}
&&{\mathbb{S}}_{\Delta} = \{(x,u^{(1)}) : \CA^i{}^{\mu}_{\alpha}
u^{\alpha}_i = 0, \quad \mu=1,\ldots, l\}\\
&&\text{and}\\
&&\mathbb{S}_{Q} = \{(x,u^{(1)}) : \xi^i_a (u) u^{\alpha}_i = 0,
\quad \alpha=1,\ldots,q, \quad a=1,\ldots,p-k\},
\end{eqnarray*}
respectively.

A vector field $X_a$ is called a conditional symmetry of the
original system (\ref{quasimatrix}) if $X_a$ is tangent to
$\mathbb{S} = \mathbb{S}_{\Delta} \cap \mathbb{S}_Q$, i.e.
\begin{equation}
  \mathrm{pr}^{(1)} X_a \Big|_{\mathbb{S}} \in
  T_{(x,u^{(1)})}\mathbb{S},
\end{equation}
where $\mathrm{pr}^{(1)} X_a$ is the first prolongation of $X_a$
defined on $J^1(X\times U)$ and is given by
\begin{equation}
  \mathrm{pr}^{(1)} X_a = X_a - \xi^i_{a, u^{\beta}} u^{\beta}_j
  u^{\alpha}_i \frac{\p}{\p u^{\alpha}_j}, \quad a=1,\ldots, p-k,
\end{equation}
and $T_{(x,u^{(1)})} \mathbb{S}$ is the tangent space to
$\mathbb{S}$ at some point $(x,u^{(1)}) \in J^1(X\times U)$.

An Abelian Lie algebra $L$ spanned by the vector fields $X_1,
\ldots, X_{p-k}$ is called a conditional symmetry algebra of the
original system (\ref{quasimatrix}) if the following condition
\begin{equation}
\label{cond-symmetry}
  \mathrm{pr}^{(1)} X_a (\CA^i u_i)\Big|_{\mathbb{S}} =0, \quad
  a=1,\ldots, p-k,
\end{equation}
is satisfied.

Note that every solution of the overdetermined system
(\ref{over-system-DC}) can be represented by its graph
$\{(x,u(x))\}$, which is a section of $J^0$.  The conditional
symmetry algebra $L$ of (\ref{quasimatrix}) defines locally the
action of the corresponding Lie group $G$ on $J^0$.  The
symmetry group $G$ transforms certain solutions of (\ref{over-system-DC})
into other solutions of (\ref{over-system-DC}).  If the graph of a solution is preserved
by $G$ then this solution is called $G$-invariant.

Assume that $L$, spanned by the vector fields $X_1, \ldots,
X_{p-k}$, is a conditional symmetry algebra of the system
(\ref{quasimatrix}).  A solution $u=f(x)$ is said to be a
conditionally invariant solution of the system (\ref{quasimatrix})
if the graph $\{(x,f(x))\}$ is invariant under the vector fields
$X_1,\ldots,X_{p-k}$.

\begin{prop}
A nondegenerate quasilinear hyperbolic system of first order PDEs
(\ref{quasimatrix}) in $p$ independent variables and $q$ dependent
variables admits a $(p-k)$-dimensional conditional symmetry algebra
$L$ if and only if $(p-k)$ linearly independent vector fields $X_1,
\ldots, X_{p-k}$ satisfy the conditions (\ref{traceeq4}) and
(\ref{traceeq5}) on some neighborhood of $(x_0,u_0)$ of
$\mathbb{S}$.  The solution of (\ref{quasimatrix}) which are
invariant under the Lie algebra $L$ are precisely rank-k solutions
of the form (\ref{implicit-equations-k}).
\end{prop}

\noindent {\bf Proof : } Let us describe the vector fields $X_a$ in the new coordinates $(\bx,\bu)$ on $\NR^p \times \NR^q$.  From
(\ref{rank-k-vf-rect}) and (\ref{cond-symmetry}) it follows that
\begin{equation}
  \mathrm{pr}^{(1)} X_a = X_a, \quad a = k+1, \ldots, p.
\end{equation}
Hence, the symmetry criterion for $G$ to be the symmetry group of
the overdetermined system (\ref{rank-k-syst-trans-inv})(or
(\ref{rank-k-syst-trans-inv-b})) requires that the vector fields
$X_a$ of $G$ satisfy
\begin{equation}
  X_a(\Delta) = 0,
\end{equation}
whenever equations (\ref{rank-k-syst-trans-inv})(or
(\ref{rank-k-syst-trans-inv-b})) hold.  Thus the symmetry criterion
applied to the invariance conditions (\ref{rank-k-inv}) is
identically equal to zero.  After applying this criterion to the system
(\ref{rank-k-syst-trans}) in new coordinates, carrying out the
differentiation and next taking into account the conditions
(\ref{traceeq4}) and (\ref{traceeq5}) we obtain the equations which
are identically satisfied.

The converse is also true.  The assumption that the system
(\ref{quasimatrix}) be nondegenerate means that it is locally
solvable and is of maximal rank at every point $(x_0, u_0^{(1)})
\in \mathbb{S}$.  Therefore \cite{O}, the infinitesimal symmetry
criterion is a necessary and sufficient condition for the existence
of the symmetry group $G$ of the overdetermined system
(\ref{over-system-DC}).  Since the vector fields $X_a$ form an
Abelian distribution on $X \times U$, it follows that, as we have already shown in this section, conditions (\ref{traceeq4}) and (\ref{traceeq5}) hold.  That ends the proof, since the solutions of the overdetermined
system (\ref{over-system-DC}) are invariant under the algebra $L$
generated by $(p-k)$ vector fields $X_1, \ldots, X_{p-k}$.  The
invariants of the group $G$ of such vector fields are provided by
the functions $\{r^1, \ldots, r^k, u^1, \ldots, u^q\}$.  So the
general rank-k solution of (\ref{quasimatrix}) takes the form
(\ref{implicit-equations-k}).
\begin{flushright} $\square$ \end{flushright}

 The expressions in equations (\ref{traceeq4}) and
(\ref{traceeq5}) lend themselves to further simplification. Let us
recall here that any $n$ by $n$ matrix $b$ is a root of the
Cayley-Hamilton polynomial
\begin{equation}
\label{C-H} \det{(\lambda {\cal I}_n -b)} = \lambda^n - \sum_{i=1}^n
p_i(b) \lambda^{n-i}.
\end{equation}
Faddeev's approach (\cite{G}, p.87)  provides a recursive method to
compute the coefficients $p_i(b)$, based on Newton's formulae
\begin{equation}
kp_k = s_k - p_1s_{k-1}- \ldots - p_{k-1}s_1, \quad s_k = \tr{(b^k)} = \sum_{i=1}^n \lambda_i^k, \quad k=1,\ldots,n.
\end{equation}
For example, one readily computes
\begin{eqnarray}
\label{fadeev-coeffs}
%\raisetag{15mm}
%\begin{split}
  p_1 &=& \tr{(b)}, \quad  p_2 = \frac{1}{2}
  [\tr{(b^2)}-(\tr{(b)})^2], \quad
  p_3 = \frac{1}{6}
  [(\tr{(b)})^3 - 3\tr{(b)}\tr{(b^2)} + 2\tr{(b^3)}],\nonumber\\
  p_4 &=& \frac{1}{24}[6\tr{(b^4)} - (\tr{(b)})^4 - 8\tr{(b)}\tr{(b^3)}  -
  3(\tr{(b^2)})^2 + 6(\tr{(b)})^2\tr{(b^2)}], \\
   \cdots \,\, p_n &=& (-1)^{n+1} \det{(b)}. \nonumber
%\end{split}
\end{eqnarray}
According to the Cayley-Hamilton theorem one has
\begin{equation}
\label{Newton-CH}
  b^n - \sum_{i=1}^n p_i(b) b^{n-i} = 0.
\end{equation}
Using the identity (\ref{Newton-CH}), we can simplify the
expressions (\ref{traceeq4}) and (\ref{traceeq5}) to some degree, depending on the
dimension of the matrix $b$.

As an illustration we present the simplest case of a $2$ by $2$
matrix, which corresponds to rank-2 solutions for $q=2$ unknown
functions. In this case, the expressions
(\ref{traceeq4}) and (\ref{traceeq5}) become
\begin{eqnarray}
\label{traceeq2x2-1}
& &\tr{\left(\CA^{\mu} \frac{\p f}{\p r} \lambda\right)} = 0, \quad \mu = 1,\ldots,l, \\
\label{traceeq2x2-2}
& &\tr{\left(\CA^{\mu} \frac{\p f}{\p r} \eta_a \frac{\p f}{\p r} \lambda \right)} = 0, \quad a=1, \ldots, p-1.
\end{eqnarray}
Combining (\ref{traceeq2x2-1}) and (\ref{traceeq2x2-2}) leads to the factorized
form
\begin{equation}
\label{traceeq2x2-3}
\tr{\left(\CA^{\mu} \frac{\p f}{\p r} \eta_a \frac{\p f}{\p r} \lambda\right)}
= \tr{\left(\CA^{\mu} \frac{\p f}{\p r} \left(\eta_a \frac{\p f}{\p r} - {\cal I}_2\tr{\left(\eta_a \frac{\p
f}{\p r}\right)}\right) \lambda\right)}.
\end{equation}
Note that for any invertible $2$ by $2$ matrices $M$ and $N$, the Cayley-Hamilton trace identity has the form
\begin{equation}
\label{C-H-2x2}
MN - {\cal I}_2\tr{(MN)} = -(N - {\cal I}_2\tr{(N)})(M - {\cal I}_2\tr{(M)}).
\end{equation}
Using the above equation, we rewrite (\ref{traceeq2x2-3}) in the equivalent form
\begin{equation}
\label{traceeq2x2-4}
\begin{split}
&\tr{\left(\CA^{\mu}\frac{\p f}{\p r} \left(\eta_a \frac{\p f}{\p r} - {\cal I}_2\tr{\left(\eta_a \frac{\p
f}{\p r}\right)}\right)\lambda\right)}\\
&= -\tr{\left(\CA^{\mu} \frac{\p f}{\p r} \left(\frac{\p f}{\p r} - {\cal I}_2\tr{
\left(\frac{\p f}{\p
r}\right)}\right)\left(\eta_a - {\cal I}_2 \tr{(\eta_a)}\right)\lambda\right)},
\end{split}
\end{equation}
where the matrices $M$ and $N$ are identified with $\eta_a$ and $\p f / \p r$, respectively.
Since we have
\begin{equation*}
N^2 - \tr{(N)}N = -{\cal I}_2 \det{(N)},
\end{equation*}
then, if $\det{\left(\p f / \p r\right)} \neq 0$, it follows
that
\begin{equation}
\label{traceeq2x2-5}
\tr{\left(\CA^{\mu}\left(\eta_a - {\cal I}_2\tr{(\eta_a)}\right)\lambda\right)} = 0, \quad \eta_a = \left(\frac{\p \lambda^A_{i_a}}{\p u^{\alpha}}\right) \in \NR^{2 \times 2}, \quad A=1,2.
\end{equation}
For a given system (\ref{quasimatrix}) (i.e. given functions $\CA^{\mu}$), the equations
(\ref{traceeq2x2-5}) form a bilinear system of $l(p-1)$ PDEs for $2(p-1)$
functions $\lambda^A_{i_a} (u)$.  Thus we have eliminated the matrix term $\p f / \p r$ in equations (\ref{traceeq2x2-2}).  This fact greatly facilitates our task.  The proposed procedure for constructing rank-2 conditionally invariant solutions of the system (\ref{quasimatrix}) consists of the following steps.\\
{\bf 1.} We first look for two linearly independent real-valued wave vectors $\lambda^1$ and $\lambda^2$ by solving the dispersion relation (\ref{xivec2}) associated with the initial system (\ref{quasimatrix}).

\noindent {\bf 2.} If such wave vectors do exist, we substitute them into PDEs (\ref{traceeq2x2-5}) and solve this system for $\lambda^A$ in terms of $u$.

\noindent {\bf 3.} Next, we substitute the most general solutions for $\lambda^1$ and $\lambda^2$ into equations (\ref{traceeq2x2-1}) and look for a solution $u=f(r^1,r^2)$ of this system.  Thus we obtain the explicit parametrization of the 2-dimensional submanifold ${\cal S} \subset U$ in terms of $r^1$ and $r^2$.\\
{\bf 4.} We suppose that $u=f(r^1,r^2)$ is the unique solution of PDEs (\ref{traceeq2x2-1}).  Then we restrict the wave vectors $\lambda^A$ to the submanifold ${\cal S} \subset U$.  That is, the functions $\lambda^A (u)$ are pulled back to ${\cal S}$ and become functions of the parameters $r^1$ and $r^2$ on ${\cal S}$.  We denote the function $f^*(\lambda^A)$ by $\lambda^A(r^1,r^2)$.\\
{\bf 5.} In this parametrization we can implicitly determine the value of the Riemann invariants for each solution $\lambda^A$ of (\ref{traceeq2x2-5})
\begin{equation}
\label{GMC-Riemann}
r^A(x, f(r^1,r^2)) = \lambda^A_i (r^1,r^2) x^i, \quad A=1,2.
\end{equation}
{\bf 6.} Finally, we suppose that the set of implicitly defined relations (\ref{implicit-equations-k}) and (\ref{GMC-Riemann}) between $u^{\alpha}$, $r^A$ and $x^i$ can be solved so that the functions $r^A$ and $u^{\alpha}$ can be given as graphs over an open subset ${\cal D} \subset X$.  Then the function
\begin{equation}
\label{2x2-explicit}
u=f(r^1(x),r^2(x))
\end{equation}
is an explicit rank-2 solution of the quasilinear hyperbolic system (\ref{quasimatrix}).  The graph of this solution is invariant under $(p-2)$ linearly independent vector fields $X_a$.

As another illustration, let us consider the rank-3 case when $q=k=3$.
Then the condition (\ref{traceeq5}) takes the following form
\begin{equation}
\label{q3k3-a}
\tr{\left(\CA^{\mu} \frac{\p f}{\p r} \left[\eta_{a_1} \frac{\p f}{\p r} \eta_{a_2} + \eta_{a_2} \frac{\p f}{\p r} \eta_{a_1} \right] \frac{\p f}{\p r} \lambda\right)} = 0, \quad \mu=1,\ldots,l,
\end{equation}
where
\begin{equation}
 \eta_{a_j} = \left(\frac{\p \lambda^A_{i_{a_j}}}{\p u^{\alpha}} \right) \in \NR^{3 \times 3}, \quad j=1,2, \quad A=1,2,3,
\end{equation}
and the expressions (\ref{threematrices}) become
\begin{equation*}
 \frac{\p f}{\p r} = \left( \frac{\p f^{\alpha}}{\p r^A}\right) \in \NR^{3 \times 3}, \quad \lambda = (\lambda^A_i(u)) \in \NR^{3 \times p}, \quad \CA^{\mu} = ({\CA^{\mu}}^i_{\alpha}) \in \NR^{p \times 3}.
\end{equation*}
We introduce the notation
\begin{equation}
 P:=\eta_{a_1} \frac{\p f}{\p r}, \quad Q:= \eta_{a_2} \frac{\p f}{\p r}.
\end{equation}
Then, combining equations (\ref{traceeq4}) and (\ref{q3k3-a}), we obtain
\begin{equation}
\label{q3k3-b} \tr{\left(\CA^{\mu} \frac{\p f}{\p r} \left[PQ - \mathcal{I}_3\tr{(PQ)} + QP - \mathcal{I}_3 \tr{(QP)} \right]\lambda\right)} = 0.
\end{equation}
If $\det{\left(\p f / \p r\right)} \neq 0$ (otherwise the case $q=3$ can be reduced to $q \leq 2$) then, similarly to the case $q=2$, we can eliminate the term $\p f / \p r$ from (\ref{traceeq4}) and (\ref{q3k3-b}).  The resulting expressions are still quite complicated.  Nevertheless, as we show in the examples to follow, our procedure makes the construction of rank-3 solutions feasible.

\section{The fluid dynamics equations}

\indent At this point, we would like to illustrate the proposed approach to
constructing rank-k solutions with the example of the fluid dynamics equations. The fluid
under consideration is assumed to be ideal, nonstationary,
isentropic and compressible.  We restrict our analysis to the
case in which the dissipative effects, like viscosity and thermal
conductivity, are negligible, and no external forces are considered.
Under the above assumptions, the classical fluid dynamics model is
governed by the system of equations in (3+1) dimensions of the form
\begin{equation}
\label{EQ1-1}
\begin{split}
&D\rho + \rho\mathrm{div} \vec{u} = 0,\\
&D\vec{u} + \rho^{-1}\nabla p = 0,\\
&D S = 0.
\end{split}
\end{equation}
Here we have used the following notation : $\rho$, $p$ and $S$ are the density, pressure and entropy of the fluid, respectively, $\vec{u} = (u^1, u^2,u^3)$ is the vector field of the fluid velocity and $D$ is the convective derivative
\begin{equation}
\label{EQ1-2}
D = \frac{\p}{\p t} + (\vec{u} \cdot \nabla).
\end{equation}
Equations (\ref{EQ1-1}) form a quasilinear hyperbolic homogeneous system of five equations in five unknown functions
$(\rho,p,\vec{u}) \in \mathbb{R}^5$.  The independent variables are denoted by $(x^{\mu}) = (t,x,y,z) \in X \subset
\mathbb{R}^4$, $\mu = 0,1,2,3$.  According to \cite{M},\cite{Ov} this system can be reduced to a hyperbolic system of four
equations in four unknowns $u = (u^{\mu}) = (a,\vec{u}) \in U \subset \mathbb{R}^4$ describing an isentropic ideal flow,
when the sound velocity $a$ is assumed to be a function of the density $\rho$ only.  In this case the state
equation of the media $p = f(\rho,S)$ is subjected to the differential constraints
\begin{equation}
\label{EQ1-4}
\nabla p = a^2(\rho)\nabla \rho, \quad d \ln {(a\rho^{-1/\kappa})} = 0,
\end{equation}
where $a^2(\rho) = \p f / \p\rho$, $\kappa = 2(\gamma-1)^{-1}$ and $\gamma$ is the adiabatic exponent of the fluid.
Under the assumptions (\ref{EQ1-4}), the fluid dynamics model (\ref{EQ1-1}) becomes
\begin{equation}
\label{isentropic-equations}
\begin{split}
&Da + \kappa^{-1}a \mathrm{div} \vec{u} = 0,\\
&D\vec{u} + \kappa a \nabla a = 0.
\end{split}
\end{equation}
The system of equations (\ref{isentropic-equations}) can be written in the equivalent matrix evolutionary form (\ref{quasievo}).
Here $n=3$ and the $4$ by $4$ matrix functions $\CA^1$,$\CA^2$ and $\CA^3$ take the form

\begin{equation}
\label{matrices}
\CA^i = \left(\begin{array}{cccc} u^i & \delta_{i1} \kappa^{-1} a & \delta_{i2} \kappa^{-1} a & \delta_{i3} \kappa^{-1} a \\ \delta_{i1} \kappa a & u^i & 0 & 0 \\ \delta_{i2} \kappa a & 0 & u^i & 0 \\ \delta_{i3} \kappa a & 0 & 0 & u^i
\end{array}\right), \quad i=1,2,3,
\end{equation}
where $\delta_{ij} = 1$ if $i=j$ and $0$ otherwise.  The largest Lie point symmetry algebra of these equations has been already investigated in
\cite{GL1}.  It constitutes a Galilean similitude algebra generated by $12$ differential operators
\begin{equation}
\label{group}
\begin{split}
&P_{\mu} = \p_{x^{\mu}}, \quad J_k = \epsilon_{kij} (x^i
\p_{x^j}+u^i \p_{u^j}), \quad K_i = t\p_{x^i} + \p_{u^i}, \quad i=1,2,3,\\
&F = t\p_t + x^i\p_{x^i}, \quad G=-t\p_t + a\p_a + u^i\p_{u^i}.
\end{split}
\end{equation}
In the particular case when the adiabatic exponent is $\gamma=5/3$,
this algebra is generated by $13$ infinitesimal differential operators,  namely the 12 operators  (\ref{group}) and a projective transformation
\begin{equation}
\label{projective}
C=t(t\p_t + x^i \p_{x^i} - a\p_a) + (x^i-x^0u^i)\p_{u^i}.
\end{equation}
Note that the algebras generated by (\ref{group}) and by (\ref{group}) with (\ref{projective}) are fibre preserving.  The classification of the subalgebras of these algebras into conjugacy classes is presented in \cite{GL1}.  Large classes of solutions of the system (\ref{isentropic-equations}), invariant and partially-invariant (with the defect structure $\delta = 1$), have been
obtained in \cite{GL2}.

The wave vector $\lambda$ can be written in the form $(\lambda_0,\vec{\lambda})$, where $\vec{\lambda} =
(\lambda_1,\lambda_2,\lambda_3)$ denotes a direction of wave
propagation and the eigenvalue $\lambda_0$ is a phase velocity of
a considered wave.  The dispersion relation for the isentropic equations
(\ref{isentropic-equations}) takes the form
\begin{equation}
\label{dispersion1}
\det{(\lambda_0(u)
{\cal I}_4 + \lambda_i(u) \CA^i(u))} = (\lambda_0 + \vec{u}\cdot\vec{\lambda})^2 [(\lambda_0 + \vec{u}\cdot\vec{\lambda})^2 - a^2\vec{\lambda}^2] = 0.
\end{equation}
Solving the dispersion equation (\ref{dispersion1}),
we obtain two types of wave vectors, namely the potential and rotational wave vectors
\begin{equation}
\label{wavevectors}
\begin{split}
&\mathrm{i)}\,\, \lambda^E = (\epsilon a+\vec{u}\cdot\vec{e},-\vec{e}), \quad \epsilon = \pm 1,\\
&\mathrm{ii)}\,\, \lambda^S = ([\vec{u},\vec{e},\vec{m}],-\vec{e} \times \vec{m}),
\quad |\vec{e}|^{\,2} = 1,
\end{split}
\end{equation}
where $\vec{e}$ and $\vec{m}$ are unit and arbitrary vectors,
respectively.  Here, the equation (\ref{wavevectors}ii) has a
multiplicity of 2.  The quantity $[\vec{u},\vec{e},\vec{m}]$ denotes
the determinant of the matrix based on these vectors, i.e.
$[\vec{u},\vec{e},\vec{m}] = \mathrm{det}(\vec{u},\vec{e},\vec{m}).$
Several classes of k-wave solutions of the isentropic system (\ref{isentropic-equations}), obtained via the GMC, are known \cite{B}, \cite{P4}. Applying the CSM to this system allows us to compare the effectiveness of the two approaches.

\section{Rank-1 solutions of the fluid dynamics equations}

\indent Analyzing the rank-1 solutions associated with the wave
vectors $\lambda^E$ and $\lambda^S$ we consider separately two
cases.

In the first case, the potential wave vectors are the nonzero multiples of
\begin{equation*}
\lambda^E=(\epsilon a+\vec{e}\cdot\vec{u},-\vec{e}),\quad |\vec{e}|^2=1,\quad \epsilon = \pm 1.
\end{equation*}
The corresponding vector fields $X_i$ and Riemann invariant $r(x,u)$
are given by
\begin{equation*}
\begin{split}
X_i = -(a+\ve\cdot\vec{u})^{-1} e_i \frac{\p}{\p t} + \frac{\p}{\p x^i}, \quad i=1,2,3, \quad
r(x,u) = (a+\vec{e}\cdot\vec{u})t-\vec{e}\cdot\vec{x}.
\end{split}
\end{equation*}
We can now consider rank-1 potential solutions,
invariant under the vector fields $\{X_1,X_2,X_3\}$.  The change of coordinates
\begin{equation}
\label{rank-1-cc}
%\begin{split}
\bt = t, \,\, \bx^1 = r(x,u), \,\, \bx^2 = x^2,\,\, \bx^3 = x^3,\,\,
\ba=a,\,\, \bu^1 = u^1, \,\, \bu^2 = u^2, \,\, \bu^3 = u^3,
%\end{split}
\end{equation}
on $\mathbb{R}^4 \times \mathbb{R}^4$ transforms the fluid dynamics equations (\ref{isentropic-equations}) into the
 system
\begin{equation}
\label{EQ4-1} \frac{\p \ba}{\p \bx^1} = \kappa^{-1} e_i \frac{\p
\bu^i}{\p \bx^1}, \quad \frac{\p \bu^i}{\p \bx^1} = \kappa e_i
\frac{\p \ba}{\p \bx^1}, \quad i=1,2,3,
\end{equation}
with the invariance conditions
$$\ba_{\bt} = \ba_{\bx^j} = 0, \quad \bu^{\alpha}_{\bt} = \bu^{\alpha}_{\bx^j}=0, \quad
j=2,3,\quad \alpha=1,2,3.$$
If the unit vector $\vec{e}$ has the form $\vec{e} = (\cos{\bu^1}\sin{\bu^2},\cos{\bu^1}\cos{\bu^2},\sin{\bu^1})$,
then the general rank-1 solution is given by
\begin{equation*}
\begin{split}
&\ba(\bt,\bx) = \kappa^{-1}\bx^1+a^0, \quad \bu^1(\bt,\bx) = -\ln{|C \cos{\bu^2}|}, \quad C \in \NR \\
&\bu^2(\bt,\bx) = \bu^2(\bx^1), \quad \bu^3(\bt,\bx) =
-\int_0^{\bu^2} \tan(\ln{|C \cos{s}|}) \cos{s}\, ds.
\end{split}
\end{equation*}
In particular, if $\vec{e}$ is a constant unit vector, then we can integrate (\ref{EQ4-1}) and the solution is defined implicitly by the equations
$$\ba(\bt,\bx) = \ba(\bx^1), \quad \bu^i(\bt,\bx) = \kappa e_i \ba(\bx^1) + C_i,\quad C_i \in \NR,
\quad i=1,2,3.$$
If we choose $\ba=A_1 \bx^1$, $A_1 \in \NR$ and $C_i = 0$, then the explicit invariant
solution has the form
\begin{equation}
\label{EQ4-2} a(t,x) = [A_1 (1+\kappa)t-1]^{-1} A_1
\vec{e}\cdot\vec{x}, \quad \vu(t,x) = [A_1(1+\kappa)t-1]^{-1} \kappa
A_1 (\ve\cdot\vx)\ve.
\end{equation}
Note that if the characteristics of one family associated with the
eigenvalue $\lambda_0 = a + \vec{e}\cdot\vec{u}$ intersect, then we
can choose a particular value of time interval $[t_0,T]$, where $T= (A_1 (1+\kappa))^{-1}$,
in order to exclude the possibility of a gradient catastrophe.
Hence, if the initial data are sufficiently small at $t=t_0$, then the solution (\ref{EQ4-2}) remains a rank-1 solution for the time $t \in [t_0,T)$, and no discontinuities (e.g. shock waves) can appear.

In the second case, we fix a rotational wave vector
\begin{equation*}
\lambda^S = ([\vec{u},\ve,\vec{m}],-\ve \times \vec{m}), \quad |\ve|^2=1,
\end{equation*}
and the corresponding vector fields (\ref{rank-k-vf}) are given by
$$X_i = [\vu,\ve,\vm]^{-1} (\ve \times \vm)_i \frac{\p}{\p t} + \frac{\p}{\p x^i},\quad i=1,2,3.$$
Hence, the Riemann invariant associated with $\lambda^S$ has the form
\begin{equation}
\label{S1-RI}
r(x,u) = [\vu,\ve,\vm]t-[\vx,\ve,\vm].
\end{equation}
After substituting (\ref{S1-RI}) into (\ref{rank-1-cc}), the change of coordinates transforms the initial system (\ref{isentropic-equations}) into the overdetermined system composed of the following equations
\begin{equation}
\label{EQrotred}
\left[\frac{\p \vu}{\p \bx^1}, \ve,\vm\right] = 0, \quad (e_i m_j - e_j m_i) \frac{\p \ba}{\p
\bx^1} = 0, \quad i \neq j = 1,2,3,
\end{equation}
and the invariance conditions
\begin{equation}
\label{rank-1-inv}
\ba_{\bt} = \ba_{\bx^j}=0, \quad \bu^{\alpha}_{\bt} = \bu^{\alpha}_{\bx^j} = 0, \quad
j=2,3, \quad \alpha=1,2,3.
\end{equation}
Hence, the sound velocity is constant
($a=a_0$).  If $\ve$ and $\vm$ are constant vectors such that $(e_1
m_2 - e_2 m_1) \neq 0$, then we can integrate the system composed of (\ref{EQrotred}) and (\ref{rank-1-inv}). The explicit solution is given by
\begin{equation}
\begin{split}
&\ba(\bt,\bx)=a_0, \quad \bu^1(\bt,\bx) = \bu^1(\bx^1),\quad \bu^2(\bt,\bx) = \bu^2(\bx^1),\\
&\bu^3(\bt,\bx) = (e_1 m_2 - e_2 m_1)^{-1} \left[C - (e_2 m_3-e_3 m_2)\bu^1 - (e_3 m_1 - e_1 m_3)\bu^2\right],
\end{split}
\end{equation}
where $\bu^1$ and $\bu^2$ are arbitrary functions of the Riemann
invariant, which takes the form
\begin{equation}
r(x,u) = Ct - [\vx,\ve,\vm].
\end{equation}
 As expected, this result coincides with the solution obtained through the GMC \cite{P4}.  The presence of arbitrary functions in the obtained solution allows us to find bounded solutions valid for all time $t>0$.  For example,
the bounded bump-type solution $\bu^i = \sech(A_{(i)}r^i)$, $i=1,2$, contains no discontinuities. 

\section{Rank-2 solutions}

The construction approach outlined in Section 3 has been applied to the isentropic flow equations
(\ref{isentropic-equations}) in order to obtain rank-2 and rank-3 solutions (the latter are presented in the next section).  In the case of rank-2 solutions, in order to facilitate computations, we assume that the directions of wave propagation $\vec{\lambda}^A$ are constant, but not their phase velocities $\lambda^A_0$.

After considering all possible combinations of the potential and rotational wave vectors ($\lambda^{E_i}$ and $\lambda^{S_i}$, respectively, $i=1,2,3$) we found eight cases compatible with the conditions (\ref{traceeq4}) and (\ref{traceeq5}), leading to eight different classes of solutions.  These solutions, in their general form, possess some degree of freedom, that is, depend on one or two arbitrary functions of one or two variables (Riemann invariants), depending on the case.  This arbitrariness allows us to change the geometrical properties of the governed fluid flow in such a way as to exclude the presence of singularities.  This fact is of a special significance here since, as is well known \cite{BL}, \cite{CF},\cite{RJ}, in most cases, even for arbitrary smooth and sufficiently small initial data at $t=t_0$, the magnitude of the first derivatives of Riemann invariants becomes unbounded in some finite time $T$; thus, solutions expressible in terms of Riemann invariants usually admit a gradient catastrophe.  Nevertheless, we have been able to demonstrate that it is still possible in these cases to construct bounded solutions and, in particular, soliton-like solutions, through the proper selection of the arbitrary function(s) appearing in the general solution.  To this purpose we submit this arbitrary function(s), say $v$, to the differential constraint in the form of the nonlinear Klein-Gordon equation
\begin{equation}
 \label{Klein-Gordon}
v_{tt} - \Delta v = c v^5, \quad c \in \NR
\end{equation}
which is known to possess rich families of soliton-type solutions
(see e.g. \cite{Ab},\cite{WGT}).  Equation (\ref{Klein-Gordon}) can be reduced to a
second order ODE for $v$ as a function of a Riemann invariant and
can very often be explicitly integrated.  The analysis of the
singularity structure of these ODEs allows us to select soliton-like
solutions for $v$ which, in turn, in many cases, lead to the same
type of rank-2 and rank-3 solutions of the system
(\ref{isentropic-equations}). Among them we have various types of
algebraic soliton-like solutions (admitting no singularity other
than poles), kinks, bumps and doubly periodic solutions which are
expressed in terms of Jacobi elliptic functions.

Below we list the obtained rank-2 solutions.  Some of the general
solutions found (both rank-2 and rank-3 in the next section)
coincide with the ones obtained earlier by means of the GMC.
Nevertheless, we list them all since we derive from them the
particular bounded solutions, which, to our knowledge, are all new.

For convenience, we denote by $(E_iE_j, E_iS_j, S_iS_j, E_iE_jE_k, \ldots, i,j,k =1,2,3)$ the solutions which result from nonlinear superpositions of rank-1 solutions associated with given wave vectors $\lambda^{E_i}$ or $\lambda^{S_i}$.  The sign ($+$ or $-$) coincides with the value of $\epsilon = \pm 1$ in equation (\ref{wavevectors}).

{\bf Case $\mathbf{(E_1E_2)}$.} We first discuss the superposition of two potential rank-1 solutions $E_i$ for which the
wave vectors have the form
\begin{equation}
\label{lambdaE1E2} \lambda^{E_i^{\pm}} = (\epsilon a+\ve^{\,i} \cdot \vu,
-\ve^{\,i}), \quad |\ve^{\,i}|^2=1, \quad i=1,2, \quad \epsilon = \pm 1.
\end{equation}
We assume that the wave vectors $\lambda^{{E_1}^{\pm}}$ and $\lambda^{{E_2}^{\pm}}$ are linearly independent.  The
corresponding vector fields (\ref{rank-k-vf}) are given by
\begin{equation}
\label{vfE1E2}
X_1 =\frac{\p}{\p x^2} - \frac{\sigma_2}{\beta_1} \frac{\p}{\p t} -
\frac{\beta_2}{\beta_1} \frac{\p}{\p x^1}, \quad X_2 = \frac{\p}{\p x^3} - \frac{\sigma_3}{\beta_1} \frac{\p}{\p t}
- \frac{\beta_3}{\beta_1} \frac{\p}{\p x^1},
\end{equation}
with
\begin{equation}
\begin{split}
\beta_i &= e_i^2(a+\ve^{\,1}\cdot\vu) - e_i^1(a+\ve^{\,2}\cdot\vu), \quad
i=1,2,3, \\
\sigma_j & = e_1^1e_j^2 - e_j^1e_1^2, \quad j=2,3, \quad  \epsilon = 1.
\end{split}
\end{equation}
The nonscattering rank-2 potential solution $(E^+_1E^+_2)$ has the
form
\begin{equation}
\label{solE1E2}
a = a_1(r^1) + a_2(r^2), \quad \vu = \kappa (a_1(r^1)\ve^{\,1} + a_2(r^2)\ve^{\,2}),
\end{equation}
where $a_1$ and $a_2$ are arbitrary functions of the Riemann invariants
\begin{equation}
\label{RIE1R2}
\begin{split}
&r^1(x,u) = (1+\kappa)a_1(r^1)t - \ve^{\,1} \cdot \vx, \quad |\ve^{\,1}|^2 = 1,\\
&r^2(x,u) = (1+\kappa)a_2(r^2)t - \ve^{\,2} \cdot \vx, \quad |\ve^{\,2}|^2 = 1,
\end{split}
\end{equation}
respectively, and the wave vectors $\ve^{\,1}$ and $\ve^{\,2}$ have to satisfy the condition
\begin{equation}
\label{E1E2angle}
\ve^{\,1} \cdot \ve^{\,2} + \kappa^{-1} = 0.
\end{equation}
Equation (\ref{E1E2angle}) holds if and only if the angle $\varphi$ between these vectors is
\begin{equation}
\label{E1E2angle-b} \cos{\varphi} = \frac{1}{2}(1-\gamma).
\end{equation}
This solution represents a Riemann double wave.  Here, the rank-1
solutions $E^+_i$, $i=1,2$, do not influence each other (they
superpose linearly).  This result coincides with the one obtained
earlier by means of the GMC \cite{P4}.

{\bf i) } In the particular case when $a_i(r^i) = -A_i r^i$, $A_i \in \NR$, $i=1,2$, the solution (\ref{solE1E2}) takes the explicit form
\begin{equation}
\label{E1E2exp}
\begin{split}
&a = \frac{A_1 \ve^{\,1} \cdot \vx}{1+(1+\kappa)A_1t} + \frac{A_2 \ve^{\,2} \cdot
\vx}{1+(1+\kappa)A_2t}\\
&\vu =  \frac{\kappa A_1 \ve^{\,1} \cdot \vx}{1+(1+\kappa) A_1t}\ve^{\,1} + \frac{\kappa A_2 \ve^{\,2} \cdot
\vx}{1+(1+\kappa) A_2t}\ve^{\,2}\\
\end{split}
\end{equation}
which admits the gradient catastrophe at the time $t = \min{\left(A_i^{-1}(1+\kappa)^{-1}\right)}$,
$i=1,2$.  Hence, some discontinuities can occur e.g. shock waves which correspond to the formation of a condensation jump from the compression waves related to $\lambda^{E^+_i}$.

{\bf ii) } The following bounded solution can be obtained using the DC (\ref{Klein-Gordon})
\begin{equation}
\label{solE1E2-2}
\begin{split}
a &= \sum_{i=1}^2 A_i r^i (1+B_i (r^i)^2)^{-1/2}, \quad A_i, B_i \in \NR, \quad B_i > 0, \\
\vu &= \kappa \left[\sum_{i=1}^2 A_i r^i (1+B_i (r^i)^2)^{-1/2} \ve^{\,i}\right],
\end{split}
\end{equation}
where the Riemann invariants are given by
\begin{equation}
r^i = \left[(1+\kappa) A_i r^i(1+B_i (r^i)^2)^{-1/2}\right]t - \ve^{\,i} \cdot \vx, \quad i=1,2.
\end{equation}
The result (\ref{solE1E2-2}) represents an algebraic kink-type solution which is bounded for $t>0$ while each $r^i$ possesses a discontinuity at time $T = (A_i(1+\kappa))^{-1}$.

{\bf Case $\mathbf{(E_1S_2)}$.} In the mixed case $(E^+_1S_2)$, we
consider the superposition of the rank-1 potential solution $E^+_1$
with the rank-1 rotational solution $S_2$ associated respectively
with the wave vectors
\begin{equation}
\label{lambdaE1S1}
\begin{split}
&\lambda^{E^+_1} = (a + \ve^{\,1} \cdot \vu, - \ve^{\,1}), \\
&\lambda^{S_2} = ([\vu,\ve^{\,2},\vm^2],-\ve^{\,2} \times \vm^2), \quad |\ve^{\,i}|^2 = 1, \quad i=1,2.\\
\end{split}
\end{equation}
The vector fields (\ref{rank-k-vf}) corresponding to the wave vectors (\ref{lambdaE1S1}) are
\begin{equation}
\label{vfE1S1}
X_1 = \frac{\p}{\p x^2} - \frac{\sigma_2}{\beta_1} \frac{\p}{\p t}
- \frac{\beta_2}{\beta_1} \frac{\p}{\p x^1}, \quad X_2 = \frac{\p}{\p x^3} - \frac{\sigma_3}{\beta_1} \frac{\p}{\p t}
- \frac{\beta_3}{\beta_1} \frac{\p}{\p x^1},
\end{equation}
where
\begin{equation}
\begin{split}
\beta_i &= -(\ve^{\,2} \times \vm^2)_i (a+\ve^{\,1}\cdot\vu) +
e_i^1[\vu,\ve^{\,2},\vm^2], \quad i=1,2,3,\\
\sigma_j &= -e^1_1(\ve^{\,2} \times \vm^2)_j + e_j^1(\ve^{\,2} \times \vm^2)_1,
\quad j=2,3.
\end{split}
\end{equation}
The invariant nonscattering rank-2 solution $(E^+_1S_2)$ has the form
\begin{equation}
\label{E1S1-sol}
a = a_1(r^1) + a_0, \quad \vu = \kappa a_1(r^1)\ve^{\,1} + \vu_2(r^2),
\end{equation}
where
\begin{equation}
\label{E1S1-cond1}
[\vu_2, \ve^{\,2},\vm^2]=C_2, \quad|\ve^{\,i}|^2 = 1,\quad i=1,2,
\end{equation}
and $a_1$ and $u^1_2$ are any differentiable functions of $r^1$ and
$r^2$, respectively, and the relation $u^3_2 (r^2) = C_1 u^1_2(r^2)$
holds.  Here, $a_0, C_1, C_2 \in \NR$ and $\vm^2$ is an arbitrary
constant vector.  The wave vector $\lambda^{S_2}$ takes the form
\begin{equation}
\label{lambdaE1S1-2}
\lambda^{S_2} = (C_2, -(e^1_1 e_3^1 + C_1(1-(e^1_1)^2)), -e_2^1(e_3^1 - C_1e^1_1), (1-(e_3^1)^2 +
C_1 e^1_1 e_3^1)).
\end{equation}
From (\ref{E1S1-sol}), (\ref{E1S1-cond1}) and (\ref{lambdaE1S1-2}), we get
\begin{equation}
\label{E1S1-cond2}
[\ve^{\,1},\ve^{\,2},\vm^2]=0,
\end{equation}
so the vector $\vec{\lambda}^{E^+_1} = -\ve^{\,1}$ is orthogonal to $\vec{\lambda}^{S_2} = -\ve^{\,2} \times \vm^{\,2}$.
Hence, the Riemann invariants are given by
\begin{eqnarray}
\label{E1S1-RI}
& &r^1 = ((1+\kappa)a_1(r^1)+C_2(C_1e^1_1 - e_3^1)^{-1})t - \ve^{\,1} \cdot \vx, \\
& &r^2 = C_2t - (e^1_1 e_3^1 + C_1(1-(e^1_1)^2)) x^1  -e_2^1(e_3^1 - C_1e^1_1) x^2 + (1-(e_3^1)^2 +
C_1 e^1_1 e_3^1) x^3. \nonumber
\end{eqnarray}
This solution represents a Riemann double wave.

{\bf i) }   An explicit form of the solution (\ref{E1S1-sol}) can be found when $\ve^{\,1} = \ve^{\,2} = (\cos{\varphi}, \sin{\varphi},0)$ and $\vm^{\,2} = (\sin{\varphi},-\cos{\varphi},C_1\sin{\varphi})$, $C_1 \in \NR$, and we choose $a_1(r^1) = A_1 r^1$, $A_1 \in \NR$.
The Riemann invariants are now given by
\begin{equation}
\begin{split}
&r^1 = \frac{C_2t + x^1C_1 \cos^2{\varphi} + x^2C_1 \cos{\varphi}\sin{\varphi}}{(C_1 \cos{\varphi})(A_1(1+\kappa)t - 1)},\\
&r^2 = C_2t - C_1x^1\sin^2{\varphi} +x^2 C_1\sin{\varphi}\cos{\varphi}
+ x^3,
\end{split}
\end{equation}
and the solution becomes
\begin{equation}
\begin{split}
&a= A_1\frac{C_2t + C_1 x^1\cos^2{\varphi}  + C_1
x^2\cos{\varphi}\sin{\varphi}}{(C_1 \cos{\varphi})(A_1(1+\kappa)t -
1)}, \quad u^3 = C_1 u^1_2(r^2),\\
&u^1 = \frac{\kappa A_1 (C_2t + C_1x^1\cos^2{\varphi} +
C_1x^2\sin{\varphi}\cos{\varphi})}{C_1(A_1(1+\kappa)t
-1)} + u^1_2(r^2),\\
&u^2 = \frac{\kappa A_1}{C_1} (C_2 \tan{\varphi} t +
C_1x^1\sin{\varphi}\cos{\varphi} + C_1x^2\sin^2{\varphi})\\
& \hspace{8mm} - \frac{C_2}{C_1 \sin{\varphi}\cos{\varphi}} - u^1_2(r^2)\cot{\varphi},
\end{split}
\end{equation}
where $u^1_2(r^2)$ is an arbitrary function of $r^2$.  Note that $a$
and $u^1$ admit the gradient
catastrophe at the time $T=(A_1(1+\kappa))^{-1}$.

{\bf ii) } Another interesting case of a conditionally invariant
solution occurs when we impose condition (\ref{Klein-Gordon})  on
the functions $a_1$ and $u^1_2$. Then the solution is bounded and
represents a solitary double wave of the type $(E^+_1S_2)$
\begin{equation}
\begin{split}
\label{E1S1-soliton}
&a = A_1(1+B_1 (r^1- 1)^2)^{-1/2} + a_0, \quad A_1, B_1, C_1 \in \NR, \quad B_1 > 0,\\
&\vu = \kappa A_1(1+B_1(r^1 - 1)^2)^{-1/2} \ve^{\,1} + (u^1_2(r^2), E_2u^1_2(r^2) + F_2, C_1 u^1_2(r^2))^T,
\end{split}
\end{equation}
where
\begin{eqnarray}
&&u^1_2(r^2) = A_2 ( 1+B_2 \cosh D_2(r^2 - 1))^{-1/2}, \quad A_2,B_2, D_2 \in \NR,\quad B_2 > 0,\\
&&E_2 = -(e^1_2(C_1e^1_1-e^1_3))^{-1}(C_1e^1_3+e^1_1)(C_1 e^1_1 -
e^1_3), \quad F_2 = C_2 (e^1_2(C_1e^1_1-e^1_3))^{-1}.\nonumber
\end{eqnarray}
The Riemann invariants take the form
\begin{eqnarray}
&&r^1 = (1+\kappa)(A_1(1+B_1 (r^1- 1)^2)^{-1/2}+ C_2(C_1 e^1_1 - e^1_3)^{-1})t- \ve^{\,1} \cdot \vx,\\
&&r^2 = C_2t - (e^1_1 e_3^1 + C_1(1-(e^1_1)^2)) x^1  -e_2^1(e_3^1 - C_1e^1_1) x^2 + (1-(e_3^1)^2 +
C_1 e^1_1 e_3^1) x^3.\nonumber
\end{eqnarray}
The solution remains bounded even though the function $r^1$ admits the gradient catastrophe at the time
$T = (1+B_1)^{3/2}\left[(1+\kappa)A_1B_1\right]^{-1}$.

 {\bf Case $\mathbf{(S_1S_2)}$ : } {\bf i) } Let us assume that
\begin{equation*}
\ve^{\,1} = (0,0,1), \quad \vm^1 = (0,1,0), \quad
\ve^{\,2} = (1,0,0), \quad \vm^2 = (0,0,1).
\end{equation*}
Then the wave vectors (\ref{wavevectors}ii) are given by
$\lambda^{S_1} = (-u^1,1,0,0)$ and $\lambda^{S_2}=(-u^2,0,1,0)$ and are linearly independent.
So we are looking for rank-2 solution $(S_1S_2)$ invariant under the vector fields
\begin{equation}
\label{vfS1S2} X_1 = \frac{\p}{\p t} + u^1 \frac{\p}{\p x^1} +
u^2\frac{\p}{\p x^2}, \quad X_2 = \frac{\p}{\p x^3}.
\end{equation}
The corresponding Riemann invariants are
\begin{equation}
\label{S1S2a-RI}
r^1(x,u) = x^1 - u^1t, \quad r^2(x,u) = x^2 - u^2t.
\end{equation}
The change of coordinates
\begin{equation}
\begin{split}
&\bt = t, \quad \bx^1 = x^1 - u^1t, \quad \bx^2 = x^2 - u^2t, \quad \bx^3 = x^3,\\
&\ba = a, \quad \bu^1 = u^1, \quad \bu^2 = u^2, \quad \bu^3 = u^3,
\end{split}
\end{equation}
transforms the system (\ref{isentropic-equations}) in this case into the equations
\begin{equation}
\label{S1S2-trans}
\begin{split}
& \frac{\p \bu^1}{\p \bx^1} + \frac{\p \bu^2}{\p \bx^2} = 0,\quad \frac{\p \bu^1}{\p \bx^1}\frac{\p \bu^2}{\p \bx^2} - \frac{\p \bu^1}{\p \bx^2}\frac{\p \bu^2}{\p \bx^1} =0, \\
& \frac{\p \ba}{\p \bx^1} = \frac{\p \ba}{\p \bx^2} = \frac{\p
\ba}{\p \bx^3} = 0,\quad \frac{\p \bu^i}{\p \bx^3}= 0, \quad
i=1,2,3.
\end{split}
\end{equation}
The solution of system (\ref{S1S2-trans}) has the form
\begin{equation}
\label{S1S2a-sol}
\ba = a_0, \quad \bu^1(\bt,\bx) = -\frac{\p \psi}{\p \bx^2}, \quad \bu^2(\bt,\bx) =
 \frac{\p \psi}{\p \bx^1}, \quad \bu^3(\bt,\bx) = \bu^3(\bx^1,\bx^2),
\end{equation}
where the function $\psi$ satisfies the homogeneous Monge-Amp\`ere
equation
\begin{equation}
\psi_{\bx^1\bx^1}\psi_{\bx^2\bx^2} - \psi_{\bx^1\bx^2}^2 = 0,
\end{equation}
and $\bu^3$ is an arbitrary function of two variables.  Note that this solution has rank 2 but it is not a Riemann double wave.

{\bf i) } The proper selection of the function $\psi$ transforms the
solution (\ref{S1S2a-sol}) into
\begin{equation}
\begin{split}
&a(t,x) = a_0, \quad u^1 = (1-n) \left(\frac{x^1 - u^1t}{x^2 - u^2t}\right)^n, \quad n \in \NZ \setminus \{1\},\\
&u^2 = -n \left(\frac{x^2-u^2t}{x^1-u^1t}\right)^{1-n}, \quad
u^3(t,x) = u^3(x^1-u^1t,x^2-u^2t).
\end{split}
\end{equation}
For $n=2$, we obtain an explicit solution of the form
\begin{equation}
\label{S1S2n2}
\begin{split}
&a=a_0, \quad u^1 = -2^{-1}t^{-2} [x^1t + (x^2)^2 \pm
  x^2((x^2)^2+4tx^1)^{1/2}], \\
&u^2 = t^{-1} [x^2 \pm ((x^2)^2 +
  4tx^1)^{1/2}], \quad u^3 = u^3(x^1 - u^1t, x^2 - u^2t).
\end{split}
\end{equation}
with a singularity at $t=0$.

{\bf ii) } Another example worth considering is the case when fluid
velocity can be decomposed as follows $\vu = \vu_1(r^1) +
\vu_2(r^2)$. Then we get the scattering nonsingular rank-2 solution
\begin{equation}
\label{S1S2b-sol}
\begin{split}
&u^1 = \frac{\left(C_1 - \lambda^1_2 u^2_1(r^1) - \lambda^1_3
u^3_1(r^1) \right)}{\lambda^1_1} -
\left(\frac{\lambda^2_3}{\lambda^2_1} \eta + \frac{\lambda^2_2}{
\lambda^2_1}\right)u^2_2(r^2) + \frac{C_2}{\lambda^2_1}, \quad C_1, C_2 \in \NR,
\\
&u^2 = u^2_1(r^1) + u^2_2(r^2), \quad u^3 = u^3_1(r^1) + \eta
u^2_2(r^2), \quad a=a_0,\quad \eta = \frac{\lambda^2_1\lambda^1_2 -
\lambda^1_1\lambda^2_2}{\lambda^1_1\lambda^2_3 -
\lambda^1_3\lambda^2_1},
\end{split}
\end{equation}
where we introduced the notation $\lambda^{S_i} = (\lambda^i_0,
\vec{\lambda}^i)$, $i=1,2$.  The above solution is invariant under
the vector fields
\begin{equation}
X_1 = \frac{\p}{\p x^2} - \frac{\sigma_2}{\beta_1} \frac{\p}{\p t} - \frac{\beta_2}{\beta_1} \frac{\p}{\p x^1},\quad
X_2 = \frac{\p}{\p x^3} - \frac{\sigma_3}{\beta_1} \frac{\p}{\p t} - \frac{\beta_3}{\beta_1} \frac{\p}{\p x^1},
\end{equation}
with
\begin{equation}
\begin{split}
\sigma_i = \lambda^1_1 \lambda^2_i - \lambda^1_i \lambda^2_1, \quad
\beta_j = \lambda^2_j[\vu,\ve^{\,1},\vm^1] - \lambda^1_j [\vu,\ve^{\,2},\vm^2],\,\, i=2,3, \,\, j=1,2,3.
\end{split}
\end{equation}
Here $u^2_1$ and $u^3_1$ are arbitrary functions of $r^1$, $u^2_2$ is an arbitrary function of $r^2$ and  $C_i = [\vu_i(r^i),\ve^{\,i},\vm^{\,i}]$, $i=1,2$. The Riemann
invariants take the form
\begin{equation}
\label{S1S2b-RI}
\begin{split}
&r^1 = (C_1 + C_2\lambda^1_1 / \lambda^2_1)t - \vec{\lambda}^1\cdot\vx,\\
&r^2 = \left(C_2 + \frac{\lambda^2_1}{\lambda^1_1} C_1 +
\left(\lambda^2_2 - \frac{\lambda^2_1
\lambda^1_2}{\lambda^1_1}\right)u^2_1(r^1) + \left(\lambda^2_3 -
\frac{\lambda^2_1\lambda^1_3} {\lambda^1_1}\right)u^3_1(r^1)\right)t
- \vec{\lambda}^2\cdot\vx.
\end{split}
\end{equation}
Note that the Riemann invariant $r^2$ depends functionally on $r^1$.  This means
that the interacting waves influence each other and superpose nonlinearly.  The result is a Riemann double wave.

{\bf iii) } By submitting the arbitrary functions $u^2_1$, $u^3_1$ and $u^2_2$ to the DC (\ref{Klein-Gordon}) we can construct the rank-2 algebraic kink-type solution of the form
\begin{equation}
\begin{split}
\label{S1S2-soliton-a}
&u^1 = (\lambda^1_1)^{-1} [ C_1 -  \lambda^1_2 A_2 r^1(1+B_2 (r^1)^2 )^{-1/2} - \lambda^1_3 A_3
r^1(1+B_3(r^1)^2)^{-1/2}]\\
& \quad - (\lambda^2_1)^{-1}[-C_2 + (\lambda^2_3 \eta + \lambda^2_2)A_1 r^2 (1+B_1 (r^2)^2)^{-1/2}], \quad A_i, B_i \in \NR, \\
&u^2 = A_1 r^2 (1+B_1 (r^2)^2)^{-1/2} + A_2 r^1 (1+B_2(r^1)^2)^{-1/2},\quad B_i > 0,\quad i=1,2,3,\\
&u^3 = A_3 r^1 (1+B_3 (r^1)^2)^{-1/2} + \eta A_1 r^2 (1+B_1(r^2)^2)^{-1/2}, \quad a = a_0,
\end{split}
\end{equation}
where the Riemann invariants are given by
\begin{equation}
\begin{split}
&r^1 = (C_1 + C_2 \frac{\lambda^1_1}{\lambda^2_1})t - \vec{\lambda}^1 \cdot \vx,\\
&r^2 = \left[C_2 + C_1 \frac{\lambda^2_1}{\lambda^1_1} + \left(\lambda^2_2 - \frac{\lambda^2_1
\lambda^1_2}{\lambda^1_1}\right) A_2 r^1 (1+B_2 (r^1)^2)^{-1/2}\right.\\
& \left.\quad\quad + \left(\lambda^2_3 - \frac{\lambda^2_1 \lambda^1_3}{\lambda^1_1} \right) A_3 r^1 (1+B_3
(r^1)^2)^{-1/2}\right]t - \vec{\lambda}^2 \cdot \vx.
\end{split}
\end{equation}

 {\bf Case $\mathbf{(E_1E_2S_3)}$ . } The nonscattering rank-2 solution
$(E^+_1E^+_2S_3)$ invariant under the vector field
\begin{equation}
X = \frac{\p}{\p x^3} - \frac{\epsilon_{ijk} \, e_i^1e_j^2(\ve^{\,3}\times\vm^3)_k}{\beta_{12}}\frac{\p}{\p t} +
\frac{\beta_{23}}{\beta_{12}} \frac{\p}{\p x^1} +
\frac{\beta_{31}}{\beta_{12}} \frac{\p}{\p x^2},
\end{equation}
with
\begin{equation}
  \begin{split}
    &\beta_{ij} = (e_j^1e_i^2 - e_i^1e_j^2)[\vu,\ve^{\,3},\vm^3]
    + (e_j^2(\ve^{\,3} \times \vm^3)_i -
    e_i^2(\ve^{\,3}\times\vm^3)_j)(a+\ve^{\,1}\cdot\vu), \\
    & \hspace{1cm} + (e_i^1(\ve^{\,3} \times \vm^3)_j -
    e_j^1(\ve^{\,3}\times\vm^3)_i)(a+\ve^{\,2}\cdot\vu), \quad i,j=1,2,3,
  \end{split}
\end{equation}
 has the form
\begin{equation}
\label{E1E2S1-sol}
\begin{split}
&a = \frac{A_1 ((e^1_1 + e_1^2)x^1 + (e_2^1 + e_2^2)x^2)}{1-A_1 (1+\kappa)t}, \quad u^3 = u^3_0,\\
&u^1 = \frac{-\kappa A_1  \left(((e_1^1)^2 + (e_1^2)^2)x^1 + (e^1_1e_2^1 + e_1^2e^2_2)x^2 \right)- u^1_3 (r^3)
}{1-A_1 (1+\kappa)t},\\
&u^2 = \kappa A_1 \left( \frac{e_2^1 \left(\beta
u^1_3(r^3)t -e^1_1x^1 -e_2^1x^2\right)}{1-A_1 (1+\kappa)t} \right.\\
& \left.\hspace{1cm} + \frac{ e^2_2\left(-\beta u^1_3(r^3)t
-e_1^2x^1 - e^2_2x^2\right)}{1-A_1 (1+\kappa)t}\right) + \frac{e^2_2 -
e_2^1}{e_1^2 - e^1_1} u^1_3(r^3),
\end{split}
\end{equation}
where $|\ve^{\,1}|^2 = |\ve^{\,2}|^2 = 1$, $\ve^{\,1}\cdot \ve^{\,2} = -\kappa^{-1}$,
$e_3^1 = e_3^2 = 0$, $\beta = (1+\kappa^{-1})/(e^1_1-e_1^2)$ and $A_1, u^3_0 \in \NR$.  The
Riemann invariants are
\begin{equation}
\label{E1E2S1-RI}
%\begin{split}
r^1 = \frac{\beta u^1_3(r^3)t - e^1_1 x^1 - e_2^1 x^2}{1 - A_1
(1+\kappa)t}, \,\, r^2 = \frac{-\beta u^1_3(r^3)t - e_1^2 x^1 -
e^2_2 x^2}{1 - A_1  (1+\kappa)t},\,\, r^3 = x^3 - u^3_0t,
%\end{split}
\end{equation}
where $u^1_3$ is an arbitrary function of $r^3$.

This solution represents a Riemann double wave.  It does not admit
removable singularities for any choice of $u^1_3(r^3)$, but the
functions $a,u^1$ and $u^2$ are subject to the gradient catastrophe
at the time $T=(A_1 (1+\kappa))^{-1}$.

{\bf Case $\mathbf{(E_1S_2S_3)}$ . } The nonscattering rank-2 solution $(E^+_1S_2S_3)$
invariant under the vector field
\begin{equation}
X = \frac{\p}{\p x^3} + \frac{\epsilon_{ijk} e_i^1 (\ve^{\,2} \times \vm^{\,2})_j (\ve^{\,3} \times \vm^{\,3})_k}{\beta_{12}}\frac{\p}{\p x^1} + \frac{\beta_{23}}{\beta_{12}}\frac{\p}{\p x^2} + \frac{\beta_{31}}{\beta_{12}}\frac{\p}{\p x^3},
\end{equation}
with
\begin{equation}
\begin{split}
&\beta_{ij} = [(\ve^{\,2} \times \vm^{\,2})_i(\ve^{\,3}\times\vm^{\,3})_j - (\ve^{\,2} \times \vm^{\,2})_j(\ve^{\,3}\times\vm^{\,3})_i](a+\ve^{\,1} \cdot \vu)\\
&\hspace{1cm} + [e_j^1(\ve^{\,3} \times \vm^{\,3})_i - e_i^1 (\ve^{\,3} \times \vm^{\,3})_j][\vu,\ve^{\,2},\vm^{\,2}] \\
&\hspace{1cm} + [e_i^1(\ve^{\,2} \times \vm^{\,2})_j - e_j^1 (\ve^{\,2} \times \vm^{\,2})_i][\vu,\ve^{\,3},\vm^{\,3}],
\end{split}
\end{equation}
is given by
\begin{equation}
\label{E1S1S2a-sol}
\begin{split}
&a =  A_1 \frac{\left(C_2/\lambda^2_1 + C_3 / \lambda^3_1\right)t - x^1}{1- A_1 (1+\kappa)t},\quad
u^1 = \frac{\left(C_2/\lambda^2_1 + C_3 / \lambda^3_1\right)(1- A_1  t) - \kappa A_1  x^1}{
1- A_1 (1+\kappa)t},\\
&u^2 = C(b \lambda^2_1 - \lambda^3_1)(\lambda^2_2 x^2 + \lambda^2_3 x^3),\quad
u^3 = -\frac{C\lambda^2_2 (b\lambda^2_1 -
\lambda^3_1)}{\lambda^2_3\lambda^3_1}(\lambda^2_2 x^2 + \lambda^2_3
x^3).
\end{split}
\end{equation}
The Riemann invariants have the explicit form
\begin{equation}
\label{E1S1S2a-RI}
\begin{split}
r^1 &= \frac{(C_2/\lambda^2_1 + C_3/\lambda^3_1)t - x^1}{1-  A_1 (1+\kappa)t},\quad A_1, C \in \NR,\\
r^2 &= \left(\kappa A_1  \frac{(C_2 + \lambda^2_1/\lambda^3_1)t -
\lambda^2_1 x^1}{1- A_1 (1+\kappa)t} + C_2 +
\frac{\lambda^2_1}{\lambda^3_1}C_3\right)t - \vec{\lambda}^2 \cdot \vx,\\
r^3 &= \left(\kappa A_1  \frac{(\lambda^3_1/\lambda^2_1 + C_3)t - \lambda^3_1 x^1}{1- A_1 (1+\kappa)t}
+ \frac{\lambda^3_1}{\lambda^2_1}C_2 + C_3\right)t - \lambda^3_1 x^1 - b(\lambda^2_2 x^2 +
\lambda^2_3 x^3),
\end{split}
\end{equation}
Here, we introduced the notation $\lambda^{S_i} = (\lambda^i_0, \vec{\lambda}^i)$, $C_i = [\vu_i, \ve^{\,i}, \vm^{\,i}]$, $i=2,3$ and $\vec{\lambda}^{\,3} = (\lambda^3_1, b\lambda^2_2, b\lambda^2_3)$, $b\in \NR$ .  Note that $a$ and $u^1$ both admit the gradient catastrophe at the time
$T=( A_1 (1+\kappa))^{-1}$ while $u^2$ and $u^3$ are stationary.  In this case the solution again has a form of Riemann double wave.

{\bf Case $\mathbf{(S_1S_2S_3)}$ . } The rank-2 solution is invariant under the vector
field
\begin{equation}
X = \frac{\p}{\p t} + u^1 \frac{\p}{\p x^1} + u^2 \frac{\p}{\p x^2} + u^3 \frac{\p}{\p x^3}.
\end{equation}
In this case subjecting the initial system (\ref{isentropic-equations}) to the DCs (\ref{rank-k-DCs}) leads to the overdetermined system
\begin{equation}
\label{isentropic-S1S2S3}
a = a_0, \quad \vu + (\vu \cdot \nabla) \vu = 0, \quad \nabla\vu = 0, \quad a_0 \in \NR.
\end{equation}
The solution of (\ref{isentropic-S1S2S3}) is divergence free if and only if
\begin{equation}
\vu = f(r^1,r^2,r^3), \quad f : \NR^3 \to \NR^3, \quad r^i = x^i-
u^it, \quad i=1,2,3.
\end{equation}
The Jacobi matrix $Df(r) = (\p f^{\alpha} / \p r^i)$ has to be
nilpotent.  In fact, the reduced system (\ref{isentropic-S1S2S3}) mandates that the
characteristic polynomial is equal to
\begin{equation}
\begin{split}
   & \det(\lambda {\cal I}_3 + Df(r)) \\
   &  \quad =\lambda^3 - \lambda^2
  \tr{(f^{\alpha}_{,r^i})} + \frac{1}{2} [(\tr{(f^{\alpha}_{,r^i})})^2 -
  \tr{(f^{\alpha}_{,r^i})^2}]\lambda + \det{(f^{\alpha}_{,r^i})} = \lambda^3.
\end{split}
\end{equation}
In order to satisfy this condition we can select the arbitrary functions $f^{\alpha}$ in the following way $f^1 = b(r^2,r^3)$ and $f^2 = f^3 = g(r^2 - r^3)$. Then we have
\begin{equation}
Df(r) = \left(\begin{array}{ccc} 0 & b_{r^2} & b_{r^3} \\ 0 & g_s &
-g_s \\ 0 & g_s & -g_s
\end{array} \right), \quad s=r^2 - r^3.
\end{equation}
If  $b_{r^2} \neq b_{r^3}$, then $\rank{ Df(r)} = 2$, otherwise $f^1$ is an arbitrary function of one variable, i.e. $f^1
= h(r^2 - r^3)$, and $\rank{Df(r)} = 1$.  In the rank-2 case the solution
has the form
\begin{equation}
\label{S1S2S3-sol}
\begin{split}
&u^1(x,t) = b(x^2 - tg(x^2 - x^3), x^3 - tg(x^2 - x^3)),\\
&u^2(x,t) = u^3(x,t) = g(x^2 - x^3), \quad a=a_0, \quad a_0 \in \NR,
\end{split}
\end{equation}
where $b$ is an arbitrary function of two variables $(x^2-u^2t)$ and $(x^3-u^3t)$,
and $g$ is an arbitrary function of $(x^2-x^3)$.

Depending on the choice of the arbitrary functions, the relations
(\ref{S1S2S3-sol}) can lead to elementary solutions (constant,
algebraic, with one or two poles, trigonometric, hyperbolic)  or
doubly periodic solutions which can be expressed in terms of the Jacobi's elliptic functions $\mathrm{sn},\mathrm{cn}$ and
$\mathrm{dn}$.  To ensure that the elliptic solutions possess one
real and one purely imaginary period and that, for real argument
$r^i$, they are contained in the interval between $-1$ and $+1$, the
moduli $k$ of the elliptic functions have to satisfy the condition
$0 < k^2 < 1$.  An example of such elliptic solution has been
obtained by submitting the arbitrary functions $b$ and $g$ to the DC
(\ref{Klein-Gordon}).  It has the explicit form
\begin{eqnarray}
\label{S1S2S3-soliton-a}
& &u^1 = A_1[1 + B_1 \mathrm{sn}^2 (\beta(x^2 + nx^3) - (n+1)[A_2 [1+B_2 \mathrm{sn}^2(\beta(x^2-x^3),k)]^{-1/2}], k) ]^{-1/2},\nonumber \\
& &u^2 = u^3 = A_2 [1+B_2 \mathrm{sn}^2(\beta(x^2-x^3),k)]^{-1/2} \mathrm{sn}(\beta(x^2-x^3),k),\\
& &a = a_0, \quad 0 < k^2 < 1, \quad  A_i, B_i, \beta \in \NR, \quad B_i >0, \quad i=1,2. \nonumber
\end{eqnarray}
This is a bounded solution representing a snoidal double wave.

\section{Rank-3 solutions}
Let us now present the rank-3 solutions obtained by way of the
procedure analogical to the one described in Section 3 for the
rank-2 solutions.

{\bf Case $\mathbf{(E_1E_2E_3)}$ .} The rank-3 potential solution
$(E^+_1E^+_2E^+_3)$ invariant under the vector field
\begin{equation}
    X =\frac{\p}{\p x^3} - \frac{[\ve^{\,1},\ve^{\,2},\ve^{\,3}]}{\beta_3} \frac{\p}{\p
    t} + \frac{\beta_1}{\beta_3} \frac{\p}{\p x^1} + \frac{\beta_2}{\beta_3} \frac{\p}{\p x^2},
\end{equation}
with
$\beta_i = (\ve^{\,2} \times \ve^{\,3})_i(a+\ve^{\,1}\cdot\vu) +
(\ve^{\,1}\times\ve^{\,3})_i(a+\ve^{\,2}\cdot\vu) +
(\ve^{\,1}\times\ve^{\,2})_i(a+\ve^{\,3}\cdot\vu),$ takes the form
\begin{equation}
\label{E1E2E3-sol1}
a = a_1(r^1) + a_2(r^2) + a_3(r^3), \quad \vu = \kappa(\ve^{\,1}a_1(r^1) + \ve^{\,2}a_2(r^2) + \ve^{\,3}a_3(r^3)),
\end{equation}
where the Riemann invariants are
\begin{equation}
r^i(x,u) = (1+\kappa) a_i(r^i)t - \ve^{\,i} \cdot \vx, \quad  \ve^{\,i} \cdot \ve^{\,j} = -\kappa^{-1}, \quad |\ve^{\,i}|^2=1, \quad i\neq j = 1,2,3,
\end{equation}
and $a_i$ are arbitrary functions of $r^i$.  Note that, just as in the
case $E_1E_2$,  the angle $\varphi_{ij}$ between each pair of wave
vectors $\ve^{\,i},\ve^{\,j}$, $i\neq j=1,2,3$, has to satisfy the
condition (\ref{E1E2angle-b}).  This nonscattering rank-3 solution
coincides with the one obtained previously by the GMC \cite{P4}.

{\bf i) } After submitting the arbitrary functions $a_i$ to the DC
(\ref{Klein-Gordon}) we obtain several bounded solutions.  We list
here two examples.

An interesting case is the algebraic kink solution
\begin{equation}
\begin{split}
\label{E1E2E3-soliton-a}
a = \sum_{i=1}^3 A_i r^i (1+B_i (r^i)^2)^{-1/2}, \quad
\vu = \kappa \left[\sum_{i=1}^3 A_i r^i (1+B_i (r^i)^2)^{-1/2} \ve^{\,i}\right],
\end{split}
\end{equation}
where the Riemann invariants are given by
\begin{equation}
r^i = \left[(1+\kappa) A_i r^i(1+B_i (r^i)^2)^{-1/2}\right]t - \ve^{\,i} \cdot \vx, \quad A_i, B_i \in \NR \quad i=1,2,3.
\end{equation}
This solution evolves as a triple wave and is bounded even when the Riemann invariants $r^i$
admit the gradient catastrophe at the time $T_i = (1+\kappa)^{-1} A_i^{-1}$.

{\bf ii) } Another interesting solution describes an algebraic solitary triple wave of a kink type
\begin{equation}
\label{E1E2E3-soliton-b}
a = \sum_{i=1}^3 A_i (1+e^{B_i r^i})^{-1/2}, \quad \vu = \kappa \sum_{i=1}^3 A_i (1+e^{B_i r^i})^{-1/2} \ve^{\,i}, \quad A_i, B_i \in \NR\\
\end{equation}
where the Riemann invariants are given by
\begin{equation}
r^i = \left[(1+\kappa) A_i (1+e^{B_i r^i})^{-1/2}\right] t - \ve^{\,i} \cdot \vx, \quad i=1,2,3.
\end{equation}
The Riemann invariants admit the gradient catastrophe at the time
\begin{equation}
T_i = -2^{5/2} ((1+\kappa)A_iB_i)^{-1},
\end{equation}
but the solution remains bounded. In both cases the angle $\varphi_{ij}$ between the wave vectors $\ve^{\,i}$ and $\ve^{\,j}$ is given by (\ref{E1E2angle-b}).

{\bf Case $\mathbf{(E_1S_2S_3)}$ : } In this case we have to
distinguish two situations, depending on the choice of wave vectors
$\lambda^{E_1}$, $\lambda^{S_2}$ and $\lambda^{S_3}$.

First we look for the rank-3 solution $(E^+_1S_2S_3)$ invariant
under the vector field
\begin{equation}
\label{E1S1S2-vf}
\begin{split}
X = e^1_2 \frac{\p}{\p x^1} + e^2_2 \frac{\p}{\p x^2},
\end{split}
\end{equation}
where we have assumed that the linearly independent wave vectors
associated with the waves $E_1^+, S_2$ and $S_3$ are given by
\begin{equation}
\label{E1S2S3-lambda} \lambda^{E^+_1} = (a+u^3,0,0,-1), \quad \lambda^{S_2} =
(e^2_2u^1 - e^1_2u^2, -e^2_2, e^1_2,0), \quad \lambda^{S_3} = (-e^3_3
u^3, 0, 0 ,e^3_3).
\end{equation}
The corresponding Riemann invariants satisfy the following relations
\begin{equation*}
\label{E1S1S2b-RI}
\begin{split}
r^1 = ((1+\kappa^{-1})f(r^1) + a_0 + u^3_0)t - x^3, \quad r^2 = t -
x^1\sin{g(r^2,r^3)} + x^2\cos{g(r^2,r^3)},
\end{split}
\end{equation*}
where $r^3$ obeys the evolutionary partial differential equation
\begin{equation}
\label{PDE-r3} \frac{\p r^3}{\p t} + (f(r^1)+u^3_0) \frac{\p r^3}{\p
x^3} = 0.
\end{equation}
The solution then takes the form
\begin{equation}
\label{E1S1S2b-sol}
\begin{split}
&a = \kappa^{-1}f(r^1) + a_0, \quad u^1 = \sin{g(r^2,r^3)},\\
&u^2 = -\cos{g(r^2,r^3)}, \quad u^3 = f(r^1) + u^3_0, \quad a_0,
u^3_0 \in \NR,
\end{split}
\end{equation}
where $f$ is an arbitrary function of $r^1$ and $g$ is an arbitrary
function of $r^2$ and $r^3$.  This scattering rank-3 solution has
been obtained earlier through the GMC \cite{P4}.

{\bf i) } If $f(r^1) = A_1 r^1 + B_1$, then the solution of
(\ref{PDE-r3}) can be integrated in a closed form
\begin{equation}
\begin{split}
&a = \kappa^{-1}(A_1 r^1 + B_1) + a_0, \quad u^1 = \sin{g(r^2, r^3)},\\
&u^2 = -\cos{g(r^2,r^3)}, \quad u^3 = A_1 r^1 + B_1 + u^3_0, \quad A_1, B_1 \in \NR,
\end{split}
\end{equation}
and the Riemann invariants are given by
\begin{equation}
\begin{split}
r^1 &= \frac{((1+\kappa^{-1})B_1 + a_0 + u^3_0)t - x^3}{1-(1+\kappa^{-1})A_1 t}, \\
r^2 &= t- x^1 \sin{g(r^2,r^3)} + x^2 \cos{g(r^2,r^3)},\\
r^3 &= \Psi\left[\frac{1}{A_1} (A_1(\kappa a_0 - u^3_0 )t + x^3 -\kappa a_0 - B_1) ((1+\kappa)A_1 t - \kappa)^{-\kappa/{\kappa+1}}\right],
\end{split}
\end{equation}
where $\Psi$ is an arbitrary function of its argument and $g$ is an arbitrary function of two variables $r^2$ and $r^3$.  This solution corresponds to a scattering Riemann triple wave.

After subjecting the arbitrary functions $f$ and $g$, appearing in
the solution (\ref{E1S1S2b-sol}), to the DC (\ref{Klein-Gordon}) we
get several bounded solutions. Below, we present two of them.

{\bf ii) } A physically interesting subcase of ($E^+_1S_2S_3$) is the solution
\begin{equation}
\label{8a1}
\begin{split}
&a = \kappa^{-1} A_1 [1+B_1(1+\cosh(C_1 r^1))]^{-1/2} +a_0,\\
&u^1 = \sin{\left[\frac{A_2 (R)^{-1/2} \tan{y}}{(B_2 + \tan^2{y})^{1/2}}\right]}, \quad A_i, B_i, C_1 \in \NR, \quad B_i >0,\quad i=1,2,\\
&u^2 = -\cos{\left[\frac{A_2 (R)^{-1/2} \tan{y}}{(B_2 + \tan^2{y})^{1/2}}\right]}, \quad \\
&u^3 = A_1 [1+B_1(1+\cosh(C_1 r^1))]^{-1/2} + u^3_0, \\
\end{split}
\end{equation}
Here we introduced the following notation $R = (r^2)^2 + (r^3)^2$ and $y = \frac{1}{2} \ln{|D_1 R|},$ with $D_1 \in \NR$.
The Riemann invariants $r^1$ and $r^2$ are
\begin{equation}
\begin{split}
&r^1 = (1+\kappa^{-1})A_1 [1+B_1(1+\cosh(C_1 r^1))]^{-1/2}t - x^3,\\
&r^2 = t- x^1 \sin{\left[\frac{A_2 (R)^{-1/2} \tan{y}}{(B_2 + \tan^2{y})^{1/2}}\right]} + x^2\cos{\left[\frac{A_2 (R)^{-1/2} \tan{y}}{(B_2 + \tan^2{y})^{1/2}}\right]}
\end{split}
\end{equation}
and $r^3$ satisfies the linear partial differential equation
\begin{equation}
\label{r3-PDE}
\frac{\p r^3}{\p t} + (A_1 [1+B_1(1+\cosh(C_1 r^1))]^{-1/2} + u^3_0)\frac{\p r^3}{\p x^3} = 0.
\end{equation}
This solution is finite everywhere except at $R=0$, but has discontinuities for $\ln{|D_1R|} = (2n+1)\pi$, $n\in \NZ$.  It remains bounded even when the Riemann invariants $r^1, r^2$ and $r^3$ tend to infinity.  Physically, this solution represents nonstationary concentric waves damped by the factor $R^{-1/2}$.

{\bf iii) } Another solution worth mentioning has the form of an algebraic solitary wave
\begin{equation}
\begin{split}
\label{E1S1S2-a-arb}
&a =\kappa^{-1} A_1 [1+B_1(1+\cosh C_1 r^1)]^{-1/2} + a_0 ,\quad B_1,C_1 >0,\\
&u^1 = \sin{(D_1[1+e^{h(r^2, r^3)}]^{-1/2})}, \quad A_1,B_1,C_1,D_1 \in \NR,\\
&u^2 = \cos{(D_1[1+e^{h(r^2,r^3)}]^{-1/2})},\\
&u^3 = A_1 [1+B_1(1+\cosh C_1 r^1)]^{-1/2} + u^3_0,
\end{split}
\end{equation}
where $h$ is an arbitrary function of $r^2$ and $r^3$.  The Riemann invariants are
\begin{equation}
\begin{split}
&r^1 = (1+\kappa^{-1})A_1 [1+B_1(1+\cosh(C_1 r^1))]^{-1/2}t - x^3, \\
&r^2 = t- x^1 \sin{D_1[1+e^{h(r^2, r^3)}]^{-1/2}} + x^2\cos{D_1[1+e^{h(r^2, r^3)}]^{-1/2}},
\end{split}
\end{equation}
and $r^3$ satisfies the partial differential equation
(\ref{r3-PDE}).

We now consider the case ($E_1^+ S_2S_3$) with a different selection of the wave vectors than assumed in (\ref{E1S2S3-lambda}), namely we choose
\begin{equation}
\begin{split}
&\lambda^{E^+_1} = (a+ e^1_1 u^1 + e^2_1 u^2, -e^1_1, -e^2_1,0), \quad |e_1|^2 =  1,\\
&\lambda^{S_2} = (u^2, 0, -1, 0), \quad \lambda^{S_3} = (-u^1, 1, 0, 0).
\end{split}
\end{equation}
This leads to scattering rank-3 solution of the form
\begin{equation}
\begin{split}
\label{E1S1S2c-sol}
&a=\kappa^{-1}f(r^1)+a_0 , \quad u^1 = \sin{f(r^1)}, \quad u^2 = -\cos{f(r^1)}, \\
&u^3 = g(r^2\cos{f(r^1)}+r^3\sin{f(r^1)}), \quad a_0 \in \NR,
\end{split}
\end{equation}
in which $g$ is an arbitrary function of one variable
$r^2\cos{f(r^1)}+r^3\sin{f(r^1)}$ .  The Riemann invariants are
\begin{equation}
\label{E1S1S2c-RI}
\begin{split}
r^1 &= (\kappa^{-1}f(r^1)+a_0)t - x^1 \cos{f(r^1)} - x^2 \sin{f(r^1)},\\
r^2 &= - t\cos{f(r^1)} - x^2, \quad r^3 = -t\sin{f(r^1)} + x^1.
\end{split}
\end{equation}
This triple wave solution
coincides with the one obtained through the GMC \cite{P4}.

{\bf iv) } As previously, we constructed particular solutions from (\ref{E1S1S2c-sol}) by requiring that the arbitrary function $f$ satisfies the DC (\ref{Klein-Gordon}).  One of the interesting examples is a periodic solution
\begin{equation}
\label{E1S1S2-2-sol}
\begin{split}
&\hspace{-4mm}a = \kappa^{-1} A_1 ( 1 - B_1 \cos{C_1 r^1})^{-1/2} + a_0,\\
&\hspace{-4mm}u^1 = \sin{A_1 ( 1 - B_1 \cos{C_1 r^1})^{-1/2}},\quad A_1,B_1,C_1 \in
\NR,\\
&\hspace{-4mm}u^2 = -\cos{A_1 ( 1 - B_1 \cos{C_1 r^1})^{-1/2}}, \quad |B_1|<1,\\
&\hspace{-4mm}u^3 = g(r^2 \cos{A_1 ( 1 - B_1 \cos{C_1 r^1})^{-1/2}} + r^3\sin{A_1 ( 1 - B_1 \cos{C_1 r^1})^{-1/2}}),
\end{split}
\end{equation}
with the Riemann invariants
\begin{equation}
\begin{split}
&r^1 = (\kappa^{-1} A_1 ( 1 - B_1 \cos{C_1 r^1})^{-1/2} + a_0)t - x^1\cos{A_1 ( 1 - B_1 \cos{C_1 r^1})^{-1/2}} \\
& \hspace{8mm}- x^2 \sin{A_1 ( 1 - B_1 \cos{C_1 r^1})^{-1/2}},\\
&r^2 = -t \cos{A_1 ( 1 - B_1 \cos{C_1 r^1})^{-1/2}} -x^2,\\
&r^3 = -t \sin{A_1 ( 1 - B_1 \cos{C_1 r^1})^{-1/2}} + x^1.
\end{split}
\end{equation}
This solution remains bounded even when the Riemann invariants admit a gradient catastrophe.

\section{Rank-k solutions of fluid dynamics equations}

 Let us now consider the isentropic flow of an ideal and compressible fluid in the case when the sound velocity depends on time $t$ only.  The
system (\ref{isentropic-equations}) in $(k+1)$ dimensions becomes
\begin{equation}
\label{rank-n-isentropic-transformed}
\begin{split}
&u_t + (u \cdot \nabla) u = 0, \\
&a_t + \kappa^{-1}a\dive{u} = 0, \quad a_{x^j} = 0, \quad j=1,
\ldots, k, \quad a>0, \quad \kappa = 2(\gamma - 1)^{-1}.
\end{split}
\end{equation}
We show that in this case our approach enables us to construct arbitrary rank solutions.

The change of coordinates on $\NR^{k+1} \times \NR^{k+1}$
\begin{equation}
\label{rank-n-coord2}
\bt=t, \quad \bx^1 = x^1-u^1t, \ldots, \bx^k = x^k - u^kt, \quad \ba = a,\quad \bu = u \in \NR^k,
\end{equation}
transforms (\ref{rank-n-isentropic-transformed}) into the system
\begin{equation}
\label{rank-n-transformed2} \frac{\p \bu}{\p \bt} = 0, \quad
\frac{\p \ba}{\p \bt} + \kappa^{-1}\ba \tr{\left(({\cal I}_k + \bt
D\bu(\bx))^{-1} D\bu(\bx)\right)} = 0, \,\frac{\p \ba}{\p \bx} = 0,
\end{equation}
where $D\bu(\bx)=\p \bu / \p \bx \in \NR^{k \times k}$ is the Jacobian matrix and $\bx = (\bx^1, \ldots, \bx^k) \in \NR^k$.  The general
solution of the conditions $\p \bu / \p \bt = 0$ and $\p \ba / \p \bx = 0$ is
\begin{equation}
\label{rank-n-sol2}
\bu(\bt,\bx) = f(\bx), \quad \ba(\bt,\bx) = \ba(\bt) > 0
\end{equation}
for any functions $f : \NR^k \to \NR^k$ and $\ba : \NR \to \NR$,
respectively.  Making use of (\ref{rank-n-sol2}) and of the trace identity
\begin{equation}
\label{rank-n-trace-identity}
\frac{\p}{\p \bt} (\ln{\det{B}}) = \tr{\left(B^{-1} \frac{\p B}{\p \bt}\right)},
\end{equation}
where $B=({\cal I}_k + \bt Df(\bx))$ and $Df(\bx) = \frac{\p}{\p
\bt}({\cal I}_k + \bt Df(\bx))$, we obtain from (\ref{rank-n-transformed2})
\begin{equation}
\label{rank-n-sol3} \frac{\p}{\p \bt} \left[\ln \left(|\ba(\bt)|^{\kappa}
\det{\left({\cal I}_k + \bt Df(\bx)\right)}\right)\right] = 0.
\end{equation}
Differentiating (\ref{rank-n-sol3}) with respect to $\bx$ gives the condition on the flow velocity
$f(\bx)$
\begin{equation}
\label{rank-n-sol4} \frac{\p^2}{\p \bx \p \bt}\left[\ln\left(\det{\left({\cal
I}_k + \bt Df(\bx)\right)}\right)\right] = 0.
\end{equation}
Consequently, we have
\begin{equation}
\label{rank-n-sol5} \det{\left({\cal I}_k + \bt Df(\bx)\right)} =
\alpha(\bx)\beta(\bt),
\end{equation}
where $\alpha$ and $\beta$ are arbitrary functions of their argument.  Evaluating
(\ref{rank-n-sol5}) at $\bt = 0$ implies $\alpha(\bx) = \beta(0)^{-1}.$  Therefore,
\begin{equation*}
\det{\left({\cal I}_k + \bt Df(\bx)\right)} =
\frac{\beta(\bt)}{\beta(0)},
\end{equation*}
and we obtain
\begin{equation}
\label{rank-n-sol6} \frac{\p}{\p \bx} \det{\left({\cal I}_k + \bt
Df(\bx)\right)} = 0.
\end{equation}
Equation (\ref{rank-n-sol6}) holds if and only if the coefficients $p_n$, $n=0, \ldots, k-1$, of the characteristic polynomial of the
matrix $Df(\bx)$ are constant.  Thus the general solution of
(\ref{rank-n-isentropic-transformed}) is
\begin{equation}
\label{rank-n-gen-sol}
\bu(\bt,\bx) = f(\bx), \quad \ba(\bt) = A_1\left(1+p_{k-1}\bt + \ldots + p_0
\bt^k\right)^{-1/\kappa}, \quad A_1 \in \NR^{+},
\end{equation}
with the Cauchy data
\begin{equation}
\label{rank-n-cauchy2}
t = 0, \quad u(0,x) = f(x), \quad a(0)=A_1.
\end{equation}
In the original coordinates $(x,u) \in \NR^p \times \NR^q$ this rank-k solution takes the form
\begin{equation}
u=f(x^1-u^1t, \ldots, x^k - u^kt), \quad a(t) = A_1(1 + p_{k-1}t + \ldots + p_0 t^k)^{-1/\kappa}.
\end{equation}
Note that the sound velocity $a$ is constant if and only if the Jacobian matrix $Df(\bx)$ is nilpotent, i.e.
$$\det{(-\lambda I_k + Df(\bx))} = (-\lambda)^k.$$

As an example let us consider the particular solution of (\ref{rank-n-isentropic-transformed}) for $k=2$.  It is invariant under the vector fields
\begin{equation*}
 X_j = \frac{\p}{\p t} + u^{j} \frac{\p}{\p x^{(j)}}, \quad j=1,2.
\end{equation*}
The requirement that the coefficients $p_n$ of the characteristic polynomial (\ref{fadeev-coeffs}) of the Jacobi matrix $Df(\bx)$ are constant means that
\begin{equation*}
 \det{(D(f(\bx)))}=B_1, \quad \tr{(Df(\bx))} = 2C_1, \quad B_1,C_1 \in \NR,
\end{equation*}
where we denote $B_1 = p_0$ and $2C_1= p_1$.  Solving the above conditions gives us the general rank-3 solution of (\ref{rank-n-isentropic-transformed}) which is implicitly defined by
\begin{equation}
\label{rank-n-2-sol}
\begin{split}
&u^1(t,x,y) = C_1(x-u^1t) + \frac{\p h}{\p r^1} (x-u^1t,y-u^2t), \\
&u^2(t,x,y) = C_1(y-u^2t) - \frac{\p h}{\p r^2} (x-u^1t,y-u^2t),\\
&a(t) = A_1((1+C_1t)^2 + B_1t^2)^{-1/\kappa}, \quad A_1 \in \NR^{+},
\end{split}
\end{equation}
where the function $h$ depends on two variables $r^1=x-u^1t$ and $r^2=y-u^2t$ and satisfies the nonhomogeneous Monge-Amp\`ere equation
\begin{equation}
h_{r^1r^1}h_{r^2r^2} - h_{r^1r^2} = B_1.
\end{equation}
Depending on the selection of particular solutions of this equation we obtain Riemann double waves or other types of rank-2 solutions of (\ref{rank-n-isentropic-transformed}).
\section{Summary remarks}

The objective of this paper was to develop a new systematic way of constructing rank-k solutions of quasilinear hyperbolic systems of first order PDEs in many dimensions.  Specifically, we have been interested in nonlinear superpositions of Riemann waves, which constitute the elementary solutions of these systems and are ubiquitous in the equations of mathematical physics. Interactions of Riemann waves are obviously present in many nonlinear physical phenomena. However there are still only a few examples of multiple rank solutions describing them in multi-dimensional systems.  Most of these solutions were obtained through the generalized method of characteristics.  The main idea behind our approach has been to look at this type of solutions from a different point of view, namely, to reformulate them in terms of symmetry group theory.

Let us now recapitulate our analysis.  We look for rank-k solutions of the system
(\ref{quasimatrix}), expressible in terms of Riemann invariants,
$u=f(r^1(x,u),\ldots, r^k(x,u))$, where $f: \NR^k \to \NR^q$.  Each
Riemann invariant is associated with a specific wave vector involved in the interaction, i.e. $r^A(x,u) = \lambda^A_i(u) x^i$,
$A=1,\ldots,k$, where $\mathrm{ker}(\lambda^A_i A^i(u)) \neq 0$. The
basic feature of these solutions is that they remain constant on
$(p-k)$-dimensional hyperplanes perpendicular to the set of linearly
independent wave vectors $\lambda^1, \ldots, \lambda^k$.  In the
context of group theory, this means that the graph $\{x,u(x)\}$ of
these solutions is invariant under all vector fields $X_a =
\xi^i_a(u) \p_{x^i}$ with $\lambda^A_i \xi^i_a = 0$ for $1 \leq a
\leq p-k$.  Then $u(x)$ is the solution of (\ref{quasimatrix}) for
some function $f$, because the set $\{r^1,\ldots, r^k, u^1, \ldots, u^q\}$
constitutes a complete set of invariants of the Abelian algebra $L$
of such vector fields.  The implicit form of these solutions leads
to major difficulties in applying the classical symmetry reduction
method to this case.  To overcome these difficulties, we rectify the
set of vector fields $X_a$ by a change of variables on $X\times U$,
choosing Riemann invariants as new independent variables.  The
initial equations (\ref{quasimatrix}) expressed in the new
coordinates, complemented by the invariance conditions for the
rectified vector fields $X_a$, form an overdetermined quasilinear
system (\ref{over-system-DC}).  Thus the solutions of this system
are invariant under the Abelian group corresponding to $L$. The
vector fields $X_a$ constitute the conditional symmetries of the
initial system (\ref{quasimatrix}).  The consistency conditions for
the overdetermined system (\ref{over-system-DC}), that is, the
necessary and sufficient conditions for the existence of conditionally
invariant solutions of (\ref{quasimatrix}), have been derived here
and they take the form of the trace conditions (\ref{traceeq4}) and
(\ref{traceeq5}).  Given these conditions, we were able to devise a
specific procedure for constructing solutions in terms of Riemann invariants.  We present it for the case of rank-2 solutions, however higher rank solutions can, in principle, be
constructed by analogy.  The computational difficulties should not be
underestimated here and in many cases additional assumptions are
needed in order to perform integration or to arrive at compact
forms of solutions.  Nevertheless, the implementation of the proposed CSM is still easier than of the GMC.  The latter imposes stronger restrictions on the wave vectors $\lambda^A$, which contribute to computational complexity as well as narrowing of the range of obtained solutions.

As the application to the
isentropic flow equations shows, our approach has proved quite
productive.  We were able to reconstruct the general rank-2 and
rank-3 solutions obtained via the GMC and to deliver several new
classes of solutions, namely in the cases $E_1S_1$, $S_1S_2$,
$E_1E_2S_1$, $E_1S_1S_2$ and $S_1S_2S_3$.  For the equations of an
isentropic flow with a sound velocity depending on time only (an
assumption which simplifies things considerably) we obtained the
arbitrary rank solution, together with the Cauchy conditions in a
closed form.

Moreover, we present a simple technique which allows us to overcome
the main weakness of solutions expressible in terms of Riemann
invariants, resulting from the fact that the first derivatives of
Riemann invariants, in most cases, tend to infinity after some
finite time.  We show that a proper selection of the arbitrary
functions appearing in the general solution can lead to bounded
solutions even in the cases when Riemann invariants admit a gradient
catastrophe.  We obtained numerous such solutions which, to our
knowledge, are all new (we include here only some of them, namely
(\ref{solE1E2-2}), (\ref{E1S1-soliton}), (\ref{S1S2-soliton-a}),
(\ref{S1S2S3-soliton-a}), (\ref{E1E2E3-soliton-a}),
(\ref{E1E2E3-soliton-b}), (\ref{8a1}), (\ref{E1S1S2-a-arb}) and
(\ref{E1S1S2-2-sol})).   Most of these solutions have a soliton-like
form and this fact is of note since the integrability properties of
soliton theories do not easily generalize to more than two
dimensions.

Our technique is applicable to a very wide class of systems, which
includes many physically meaningful models.  Given the promising results obtained,
we expect it may be useful in such areas as nonlinear field equations, general relativity or equations of continuous media.  Let us note also that, though the notion of Riemann invariants was originally defined for hyperbolic systems only, it seems that it can be easily adapted to elliptic systems.  Since the conditional symmetry method can be applied to these systems, it is worth investigating whether our approach to constructing rank-k solutions can be extended to the elliptic case.  Some preliminary analysis suggests that to be feasible.\\

\noindent {\bf Acknowledgement.}
%The authors thank J. Tafel (Warsaw University) for helpful and interesting
%discussions on the topic of the paper.
This work has been partially supported by research grants from
NSERC of Canada and FQRNT of Qu\'ebec.

\bibliographystyle{plain}

\end{document}